\documentclass{aa}
\usepackage{tikz}
\usetikzlibrary{shapes.geometric, arrows, positioning}
\usepackage{graphicx}
\usepackage{comment}
\usepackage{txfonts}
\usepackage{lscape}
\usepackage{subfigure}
\usepackage{natbib}
\usepackage{longtable}
\usepackage{booktabs}
\usepackage[normalem]{ulem} 
\usepackage{epstopdf}
\usepackage{xcolor}
\usepackage{soul}
\usepackage{makecell}
\usepackage{multicol,tabularx,capt-of}

\usepackage{scalerel}
\usepackage{tikz}
\usetikzlibrary{svg.path}

\defcitealias{Sanchez-Saez2023}{SS23}

\definecolor{orcidlogocol}{HTML}{A6CE39}
\tikzset{
  orcidlogo/.pic={
    \fill[orcidlogocol] svg{M256,128c0,70.7-57.3,128-128,128C57.3,256,0,198.7,0,128C0,57.3,57.3,0,128,0C198.7,0,256,57.3,256,128z};
    \fill[white] svg{M86.3,186.2H70.9V79.1h15.4v48.4V186.2z}
                 svg{M108.9,79.1h41.6c39.6,0,57,28.3,57,53.6c0,27.5-21.5,53.6-56.8,53.6h-41.8V79.1z M124.3,172.4h24.5c34.9,0,42.9-26.5,42.9-39.7c0-21.5-13.7-39.7-43.7-39.7h-23.7V172.4z}
                 svg{M88.7,56.8c0,5.5-4.5,10.1-10.1,10.1c-5.6,0-10.1-4.6-10.1-10.1c0-5.6,4.5-10.1,10.1-10.1C84.2,46.7,88.7,51.3,88.7,56.8z};
  }
}

\newcommand\orcidicon[1]{\href{https://orcid.org/#1}{\mbox{\scalerel*{
\begin{tikzpicture}[yscale=-1,transform shape]
\pic{orcidlogo};
\end{tikzpicture}
}{|}}}}

\newcommand{\xd}{X-Detected}
\newcommand{\xu}{X-Undetected}

\usepackage{hyperref} 
\hypersetup{
    colorlinks=true,
    citecolor=blue,
    linkcolor=blue,
    urlcolor=blue,
    }
\makeatletter
\renewcommand*\aa@pageof{, page \thepage{} of \pageref*{LastPage}}
\makeatother

\bibpunct{(}{)}{;}{a}{}{,} 
\usepackage{amstext} 
\definecolor{cadmiumred}{rgb}{0.89, 0.0, 0.13}

\definecolor{ste}{rgb}{0., 0.26, 0.15}

\begin{document} 
   \title{The X-ray-Weak Tail of the Quasar Population: Physical Properties and Redshift Evolution of Variability-Selected AGN in eFEDS}

   \author{ P. Ar\'evalo\inst{1,2\orcidicon{0000-0001-5675-6323}}, J. Buchner\inst{3}, M. Salvato\inst{3}, P. Sánchez-Sáez\inst{4\orcidicon{0000-0003-0820-4692}}, F. Bauer\inst{5\orcidicon{0000-0002-8686-8737}}, M.L. Martínez-Aldama\inst{6,2\orcidicon{0000-0002-7843-7689}},  D. Tubín-Arenas\inst{7\orcidicon{0000-0002-2688-7960}}, J. Cerón-Meneses\inst{1,2\orcidicon{0009-0005-3157-9435}}, R. Pucha\inst{8\orcidicon{0000-0002-4940-3009}}, P. Lira\inst{9}, A. Merloni\inst{3}, M. Mezcua\inst{10,11}, M. Krumpe\inst{7}}

   \titlerunning{X-ray properties of variability-selected AGN}
   \authorrunning{Ar\'evalo et al.}

\institute{
Instituto de F\'isica y Astronom\'ia, Facultad de Ciencias, Universidad de Valpara\'iso, Gran Breta\~na 1111, Valpara\'iso, Chile
\\e-mail: patricia.arevalo@uv.cl
\and
Millennium Nucleus on Transversal Research and Technology to Explore Supermassive Black Holes (TITANS)
\and
Max Planck Institute for Extraterrestrial Physics, Gie{\ss}enbachstra{\ss}e 1, 85748 Garching, Germany
\and
European Southern Observatory, Karl-Schwarzschild-Strasse 2, 85748 Garching bei München, Germany
\and 
Instituto de Alta Investigaci{\'{o}}n, Universidad de Tarapac{\'{a}}, Casilla 7D, Arica, 1010069, Chile
\and Astronomy Department, Universidad de Concepción, Casilla 160-C,  4030000, Concepción, Chile
\and Leibniz-Institut f\"ur Astrophysik Potsdam (AIP), An der Sternwarte 16, 14482 Potsdam, Germany
\and
Department of Physics and Astronomy, The University of Utah, 115 South 1400 East, Salt Lake City, UT 84112, USA
\and 
Departamento de Astronom\'ia, Universidad de Chile, Casilla 36D, Santiago, Chile
\and
 Institute of Space Sciences (ICE, CSIC), Campus UAB, Carrer de Can Magrans, s/n, 08193 Barcelona, Spain
\and
Institut d’Estudis Espacials de Catalunya (IEEC), Edifici RDIT, Campus UPC, 08860 Castelldefels, Barcelona, Spain
}

   \date{}
  \abstract
   {}
   {We present a comprehensive characterization of variability-selected Active Galactic Nuclei (AGN) candidates within the eROSITA Final Equatorial-Depth Survey (eFEDS), leveraging the synergy between optical variability and deep X-ray observations by SRG (Spectrum Roentgen Gamma)/eROSITA, with the aim to compare the physical properties of AGN populations selected through complementary methods.}
   {We used a public sample of AGN candidates selected via optical variability reported by Arévalo et al., restricted to the eFEDS field and cross-match it to the X-ray selected eFEDS AGN sample. Spectroscopic classifications and characterizations are obtained for the matched samples from the Dark Energy Spectroscopic Instrument public data products. 
}
   { We find a high classification purity of the variability-selected sample, with 98\% confirmed as broad-line AGN/QSOs. We identify a significant population (approximately 25\%) of variability-selected AGN that lack X-ray counterparts (\xu) in the eFEDS survey, although about 10\% of these are detected later in subsequent, shallower eROSITA scans. We find that while both \xd{} and \xu{} samples show high AGN confirmation probabilities, they exhibit distinct physical properties and redshift distributions. The \xu{} sample peaks at a higher redshift ($z \approx 2.0$) compared to the \xd{} sample ($z \approx 1.0$), with nearly half of the variability-selected AGN at $z > 2.5$ remaining undetected in X-rays. This missing fraction can be attributed to K-correction effects and SED evolution with luminosity, mainly resulting in high accretion rate objects falling out of the \xd{} sample at higher redshifts. X-ray variability accounts for a part of the missing fraction but its dependence with redshift is less clear. Ensemble spectral analysis reveals that \xd{} sources have slightly steeper optical continua and stronger (higher EW) broad emission lines (Mg\,{\sc ii}, C\,{\sc iv}, Balmer series) than their \xu{} counterparts, while the forbidden and semi-forbidden lines have equal EWs in both samples. Finally, broad absorption line (BAL) features are found almost exclusively in the \xu{} population. These results suggest that the "missing" X-ray population must be carefully accounted for when extrapolating X-ray-selected AGN to the broader Type I population, especially at high redshifts, and could impact cosmic estimates of massive black hole growth.
}
   {}

   \keywords{galaxies: active -- surveys -- methods: statistical -- methods: data analysis }
   \maketitle

\section{Introduction}\label{section:intro}

 An accurate measurement of fundamental metrics of black hole growth, such as the quasar luminosity function and its evolution, requires quasar samples that are complete and unbiased against subpopulations. This quest has driven the development of different and complementary active galactic nuclei (AGN) selection techniques. Historically, quasar surveys have relied on optical colour selection, pioneered by \citet{Sandage1965} and expanded and improved by several groups. These methods were applied systematically to increasingly larger datasets such as the Large Bright Quasar Survey \citep{Hewett95}, Palomar-Green Survey \citep{Green86}, the Hamburg/ESO Survey \citep{Wisotzki00}, the 2dF QSO Redshift Survey \citep{Croom01}, and subsequently expanded to unprecedented scale and uniformity by the Sloan Digital Sky Survey (SDSS), which used $ugriz$ photometry and colours, together with radio detections to compile the largest quasar candidate catalogue of its era \citep{Richards02}. 
 
 Colour selection methods proved to be highly efficient for bright, unobscured objects but suffer from significant incompleteness due to orientation effects, dust reddening, redshift, and host galaxy contamination in low power sources. To overcome the bias produced by dust obscuration, near-infrared (NIR) and mid-infrared (MIR) properties began to be taken into account \citep[e.g.,][]{Stern05} to successfully select even obscured quasi-stellar objects (QSOs). 
 
 As large area optical variability surveys became available, it was also viable to select quasars based on their variability properties, sometimes combined with optical and MIR colours  \citep[e.g.][]{Butler11,PalanqueDelabrouille11,Choi14,DeCicco15,Peters15,Tie17,Sanchez-Saez2021}. \citet{Sanchez-Saez19} showed that this variability selection could include quasars with redder colours, that would have been missed by colour cuts. 
 
 An alternative approach to quasar selection uses the fact that many active galaxies are X-ray emitters \citep[e.g.,][]{Avni1986, Gibson2008}, so point source X-ray detections became an excellent method to select even moderately absorbed, and host galaxy dominated AGN \citep[e.g.,][and references therein]{Brandt2015}. However, the X-ray selection is not free from biases as some populations of AGN can be intrinsically X-ray weak \citep[e.g.,][]{Brandt2000,Pu2020}. 
 
In this work, we compare the populations of AGN that can be recovered with different selection methods, in this case optical variability and X-ray surveys. The variability-selected AGN candidates were obtained with a random forest algorithm using variability features obtained from the Zwicky Transient Facility \citep[ZTF;][]{Bellm14, Bellm19}, in the optical $g$ band only, combined with single epoch photometry in other optical, NIR and MIR bands, as presented in \citet{Arevalo2025}. For the present analysis we will explore the X-ray detection rate of these variability-selected AGN and compare the properties of those with and without X-ray detections. To make a meaningful comparison, we restrict the analysis to the eFEDS \citep{Brunner2022} field, which has the great advantage of having deep, homogeneous and contiguous X-ray coverage, over 140 deg$^2$ with the eROSITA telescopes \citep{Predehl2021,Sunyaev2021}. The Dark Energy Spectroscopic Instrument \citep[DESI;][]{Levi2013,DESI2016a,DESI2016b} provides spectra and characterization of spectral features for many sources in this area. We will use these DESI data to establish the purity of the variability-based classifications and to compare the ensemble properties of the X-ray detected and X-ray undetected samples. We note that the ZTF light curves used for the classification cover the period March 2018 through July 2022, while the eFEDS X-ray observations were carried out in December 2019 i.e. during the same period. The DESI spectra used here were obtained between May 2021 and June 2022, coinciding with the second half of the ZTF lightcurves.

\section{Sample and data}\label{section:data}

Our parent sample of variability-selected AGN candidates is described in \citet{Arevalo2025}. Briefly, this sample was built using using the hierarchical random forest classifier strategy introduced in \citet{Sanchez-Saez2023}, that splits the objects into 17 different classes, including low-z, mid-z and high-z AGN and blazars. The selection was based on tens of variability features extracted from custom-made photometry on template-subtracted images provided by ZTF, for over a hundred epochs spanning 4 years (from 2018 to 2021), only in the $g$ filter. These features were complemented by catalogue optical and IR colours obtained from PanSTARRS \citep{Chambers16} and catWISE \citep{Eisenhardt2020,Marocco2021} and proper motion information provided by Gaia DR3 \citep{Gaia2023}. All these variability and colour features were used together for the classification. This classification produced 3872 AGN candidates of different classes inside the area where eFEDS reaches at least 90\% of its nominal depth \citep{Brunner2022}. This sample is mainly limited by the single-epoch depth of ZTF, leading to a g-band magnitude distribution that peaks at about $g=20.8$ and drops sharply towards dimmer objects. 

Cross-matching this sample to the eFEDS X-ray selected AGN described in \citet{Liu2022}, based on the coordinates of the counterparts identified by \citet{Salvato2022}, produces 2745 matches, leaving 1127 variability-selected AGN without an X-ray-counterpart match.

X-ray variability could preclude some sources from being detected during the eFEDS observations, which were carried out over the course of approximately two weeks. To test this hypothesis we looked for additional X-ray observations from the four independent passes that eROSITA carried out over this field as part of its main All Sky Survey \citep[eRASS,][]{Merloni2024}. We found X-ray detections for 141 of the un-matched sources, some of which had detections in more than one pass. These 141 sources are therefore considered X-ray detected. 

We note, however, that in some cases there are a few optical galaxies within the positional uncertainty of the X-ray sources. If the most likely optical counterpart is different to the variability-selected AGN in the same area, then the latter will be considered in the X-ray undetected sample. Therefore, to separate cleanly the variability-selected sources with and without X-ray counterparts, we remove from the X-ray undetected sample 32 variability-selected AGN that do have an X-ray source within 3 times the X-ray positional uncertainty. We note that 15 of these 32 objects were detected in the single passes described above and thus are moved into the X-ray detected sample. We discard the remaining 17 objects  from the analyses described below (i.e. they will not be considered in either the X-ray detected or X-ray undetected samples), unless stated otherwise.

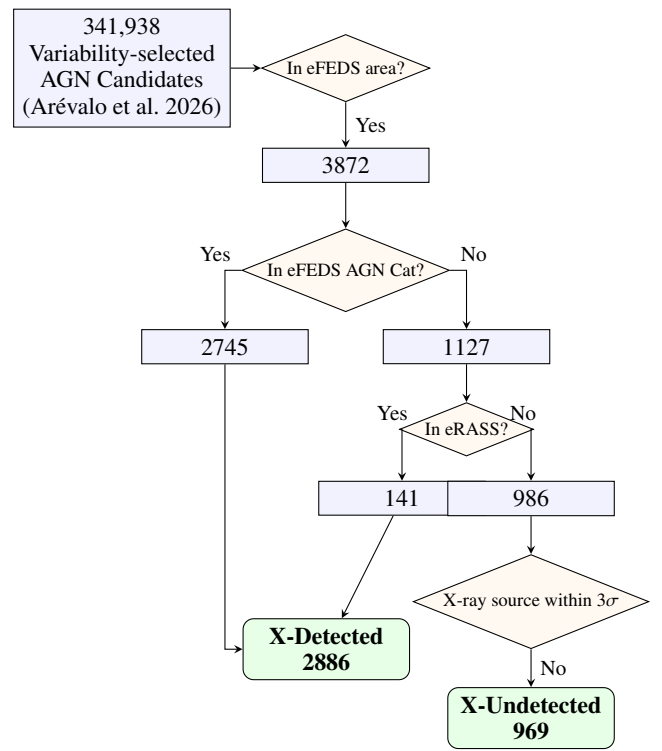
\begin{figure}[ht]
\centering
\begin{tikzpicture}[
    node distance=0.6cm and 0.4cm,
    base/.style={draw, rectangle, align=center, minimum width=2.2cm, fill=blue!5, font=\footnotesize},
    decision/.style={draw, diamond, aspect=2.5, align=center, fill=orange!5, inner sep=1pt, font=\scriptsize},
    sample/.style={draw, rectangle, rounded corners, align=center, fill=green!10, font=\footnotesize\bfseries, minimum height=0.8cm, minimum width=2.2cm}
]

\node (start) [base] {341,938 \\ Variability-selected \\ AGN Candidates \\ (Arévalo et al. 2026)};
\node (efeds_area) [decision, right=of start] {In eFEDS area?};
\node (in_area) [base, below=of efeds_area] {3872};
\node (in_catalog) [decision, below=of in_area] {In eFEDS AGN Cat?};

\node (yes_cat) [base, below left=0.5cm and -0.2cm of in_catalog] {2745};
\node (no_cat) [base, below right=0.5cm and -0.2cm of in_catalog] {1127};

\node (eRASS) [decision, below=0.5cm of no_cat] {In eRASS?};
\node (eRASS_yes) [base, below left=0.5cm and -0.7cm of eRASS] {141};
\node (eRASS_no) [base, below right=0.5cm and -0.7cm of eRASS] {986};

\node (nearby) [decision, below=0.5cm of eRASS_no] {X-ray source within 3$\sigma$};

\node (xray_detected) [sample, at={(yes_cat.center -| eRASS_yes.center)}, yshift=-4.0cm, xshift=-1.0cm] {\xd{}\\ 2886};
\node (xray_undetected) [sample, below=0.5cm of nearby] {\xu{} \\ 969};

\draw [->, >=stealth] (start) -- (efeds_area);
\draw [->, >=stealth] (efeds_area) -- node[right, font=\tiny] {Yes} (in_area);
\draw [->, >=stealth] (in_area) -- (in_catalog);

\draw [->, >=stealth] (in_catalog) -| node[above left, font=\tiny, xshift=0.2cm] {Yes} (yes_cat);
\draw [->, >=stealth] (in_catalog) -| node[above right, font=\tiny, xshift=-0.2cm] {No} (no_cat);

\draw [->, >=stealth] (no_cat) -- (eRASS);
\draw [->, >=stealth] (eRASS) -| node[above left, font=\tiny, xshift=0.2cm] {Yes} (eRASS_yes);
\draw [->, >=stealth] (eRASS) -| node[above right, font=\tiny, xshift=-0.4cm] {No} (eRASS_no);

\draw [->, >=stealth] (eRASS_no) -- (nearby);
\draw [->, >=stealth] (nearby) -- node[right, font=\tiny] {No} (xray_undetected);

\draw [->, >=stealth] (yes_cat) |- (xray_detected);
\draw [->, >=stealth] (eRASS_yes) -- (xray_detected);

\end{tikzpicture}
\caption{Flowchart of the selection of the \xd{} and \xu{}  samples. }
\label{fig:flowchart}
\end{figure}

These samples of variability-selected AGN in the eFEDS field with and without X-ray detections will be referred to as X-Detected (i.e. 2745 + 141 = 2886 objects) and X-Undetected (1127 - 141 - 17 = 969 objects) AGN candidates, respectively. Figure \ref{fig:flowchart} shows a schematic of the sample construction described above. 

We characterize below the \xd{} and \xu{} objects, estimating the purity of the AGN selection and the differences and similarities of both samples, by comparing optical spectra from the DESI first data release \citep[DR1;][]{desi_dr1}. DESI DR1 provides optical spectra, along with redshift measurements and pipeline classifications, for a large number of sources~\citep[][Bailey et al. in preparation]{desi_spec_pipeline, redrock_qso}.  Notably, the DESI spectra did not specifically target either eFEDS AGN candidates or variability-selected AGN, so the availability of spectra should not strongly bias our results. 

\section{Results}\label{section:results}
We cross-matched the \xd{} and \xu{} samples to the DESI DR1 redshift catalog compiled using the Redrock pipeline~\citep[][Bailey et al. in preparation]{redrock_qso}, namely zall-pix-iron.fits,\footnote{see \url{https://data.desi.lbl.gov/doc/organization/}} using the ZTF coordinates and the spectroscopic target coordinates, with a 1.5 arcsec matching radius. This procedure resulted in 
2760 matches for the
2886 X-Detected AGN and 844
matches for the 969
X-Undetected AGN. After removing lower quality spectra by keeping those with keywords {\tt ZWARN}=0, {\tt COADD-FIBERSTATUS}=0 and {\tt DELTACHI2}>25 leaves 
2744 spectra in the X-Detected sample and 830
spectra in X-Undetected AGN sample. 

Table \ref{tab:numbers_DESI_class} summarizes the numbers in each sample, classified as either {\tt QSO}, {\tt GALAXY} or {\tt STAR} by the DESI pipeline. The majority of variability-selected AGN are confirmed spectroscopically as QSO, whether they are detected in the X-ray band or not, although the \xu{} have a higher fraction of contaminants such as stars or galaxies.  The sources classified as {\tt GALAXY} are discussed separately in Sec. \ref{sec:varAGN_galaxy}, while objects classified as {\tt STAR} are discarded from further analysis. 

\citet{Aydar2025} published a curated, visually inspected spectroscopic redshift catalogue of the eFEDS AGN, using spectra from the Sloan Digital Sky Survey (SDSS)-V. Of the 2620 objects in the \xd{} sample that are detected in eFEDS and that have an entry in the DESI spectroscopic catalogue, 2258 have a redshift estimate in the catalogue of \citet{Aydar2025}. These values are highly consistent, with 98\% of the objects having redshift estimates within 0.05 of each other. Given this similarity and that a curated catalogue for the \xu{} sources is not available, we use the DESI redshifts for all objects, for consistency and simplicity. 

We note that our selection is based on the detections in the ZTF $g-$band filter, which has most of its transmission in the range $4000 \AA <\lambda<5500 \AA$. Therefore, the Ly$\alpha$ forest enters the filter at about $z=2.5$ and completely covers the filters wavelength range by $z=3.5$, explaining why the redshift distribution drops towards these redshifts, with only a couple of objects appearing above $z=3.6$. 

\begin{table}[]
    \centering
    \caption{Number of sources per sample and spectral classification.}
    \begin{tabular}{l|c|c}
        &X-Detected & X-Undetected \\
        \hline
        Total candidates & 2886& 969 \\
        DESI good spectra& 2744& 830\\
        DESI {\tt QSO} &2677& 760  \\
        DESI {\tt GALAXY} & 65& 40 \\
        DESI {\tt STAR} & 2 & 30 \\
        
    \end{tabular}
    \label{tab:numbers_DESI_class}
\end{table}

\begin{figure}
    \centering
    \includegraphics[width=0.99\linewidth]{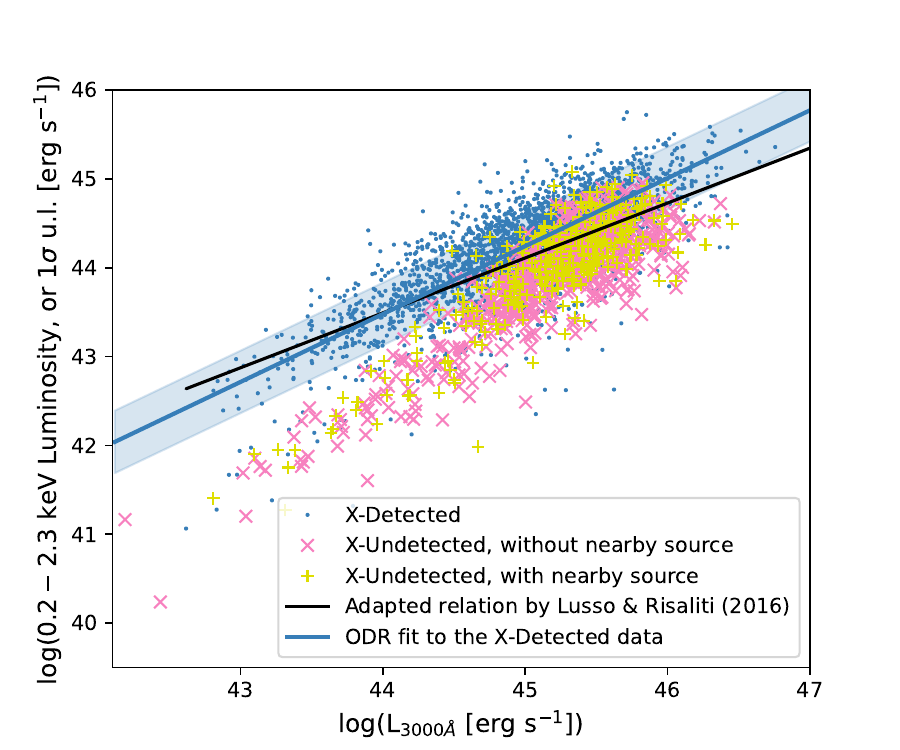}
    \includegraphics[width=0.99\linewidth]{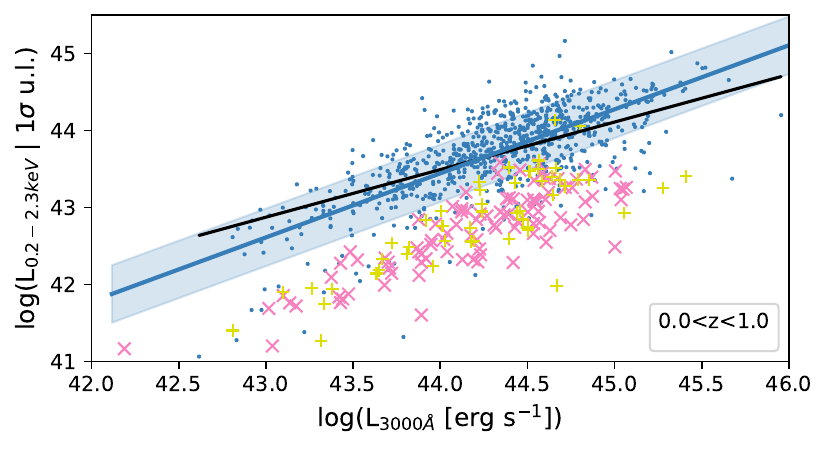}
    \includegraphics[width=0.99\linewidth]{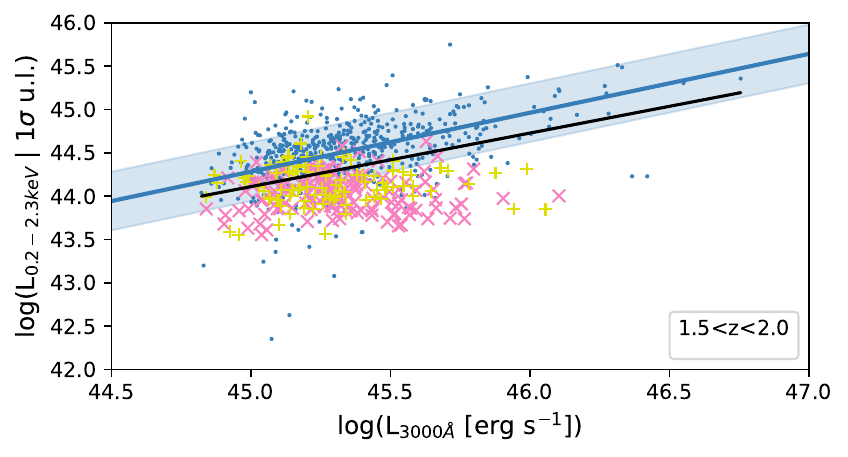}
    \includegraphics[width=0.99\linewidth]{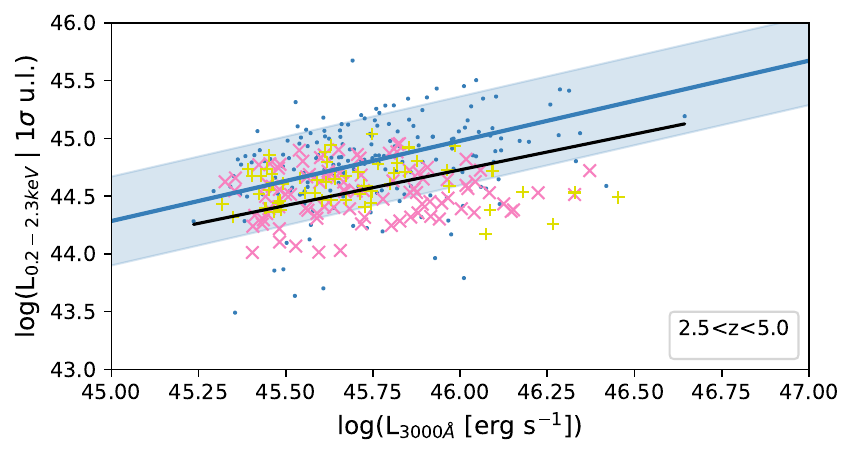}

    \caption{Rest-frame UV luminosity $L_{3000\AA}$ reported by DESI vs X-ray luminosity or 1-$\sigma$ upper limits in the rest-frame 0.2-2.3 keV range for the \xd{} sources in blue dots; variability-selected AGN that do not match an eFEDS AGN counterpart but do have an X-ray source nearby (yellow plus signs), and those that do not have a nearby X-ray source (pink crosses). A linear fit to the \xd{} sources is shown by the blue line, together with the rms of the deviations from the best-fitting model, in the shaded blue area. Black lines in all panels shows an adaptation of the relation of \citet{Lusso2016}.}
    \label{fig:upper_limits_luminosity}
\end{figure}

Figure \ref{fig:upper_limits_luminosity} displays the relation between UV and X-ray luminosities of the variability-selected AGN. For the objects without counterparts in eFEDS, we plot the 1-$\sigma$ upper limits of the eFEDS X-ray flux, obtained following the procedure described in \citet{Tubin2024}, separating those with a nearby X-ray source from those without. We note that this algorithm considers close X-ray sources as those that might potentially contaminate the X-ray upper limit, which can be tens of arcminutes away. For this plot, luminosity distances have been calculated using the redshifts provided by DESI, and an X-ray K-correction was applied considering a photon index of 1.9, while the rest-frame optical $L_{3000}$ integrated luminosity was obtained directly from the DESI FastSpecFit value-added catalogue~\footnote{\url{https://data.desi.lbl.gov/doc/releases/dr1/vac/fastspecfit/}} \citep[][Moustakas et al. in preparation]{Moustakas2023}.  

The \xd{} sample shows a good correlation between optical and X-ray luminosities. Fitting a linear relation between the logarithm of the luminosities produces a slope of $0.76 \pm 0.01$, and an rms scatter of $L_X$ data points around this model is 0.35 dex. This model is plotted as a blue line in the top panel, and the shaded area around it represents the $\pm$ rms scatter.  The X-ray upper limits fall well below this scatter band for the low luminosity objects (log $L_{3000\AA}<44.5$), while they overlap for higher luminosities. The following three panel show subsamples of the same data, split into three redshift bins, with their corresponding models in blue. At $z<1$, the upper limits mostly fall below the blue band, while in the higher redshift bins the upper limits overlap. This overlap however can be a product of the limiting depth of the X-ray data, since at each redshift bin the upper limits fall below the shaded area at the highest luminosities.

In all panels we plot the $L_{UV}$ vs $L_X$ relation found by \citet{Lusso2016}, converted to the wavelengths and units of these plots. We note the good overall correspondence in normalization and slope with our data. Our fits to the X-ray detected sources produce a steeper relation only in the lowest redshift bin, with $z<1$ (i.e. $0.83 \pm 0.03$ for our data vs 0.618 for the whole sample in \citet{Lusso2016}), while for all other redshift bins the slopes are consistent.

\section{Comparison of the \xd{} and \xu{} samples}\label{section:discussion}
In this section we explore the differences between the \xd{} and \xu{} samples, mainly using the spectroscopic results from DESI, and discuss possible explanations for the \xu{} AGN. We quantify the difference between properties of these samples mainly using information from the DESI DR1 Redrock pipeline classifications and redshifts, the fits to the emission lines and continuum published by DESI obtained using the FastSpecFit routine \citep[][Moustakas et al. in preparation]{Moustakas2023}, and the detailed classifications of the AGN/Galaxy Classification value-added catalogue\footnote{\url{https://data.desi.lbl.gov/doc/releases/dr1/vac/agngal/}} (AGN/Galaxy Classification VAC, Juneau et al. in prep.).

\subsection{Redshift distributions}
\label{sec:redshift}
\begin{figure}[htbp] 
    \centering
    \includegraphics[width=0.99\linewidth]{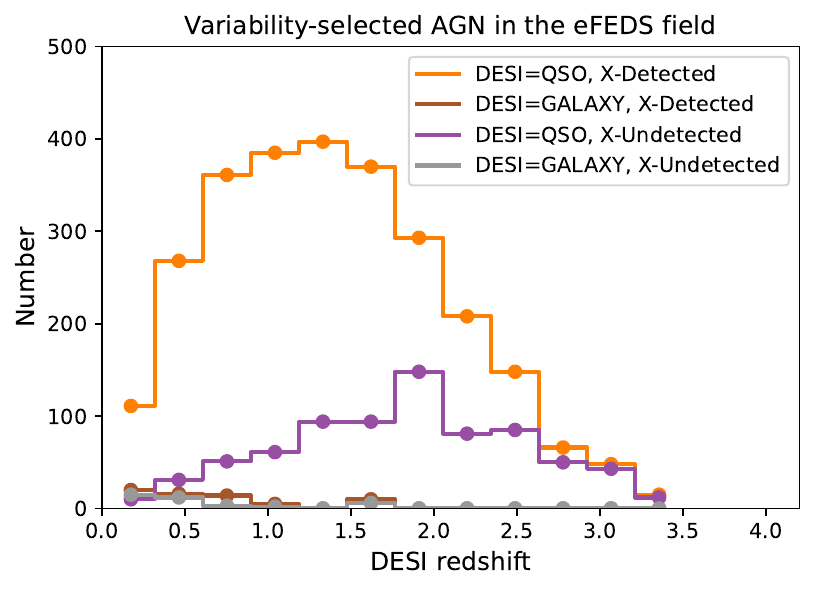}
    \includegraphics[width=0.99\linewidth]{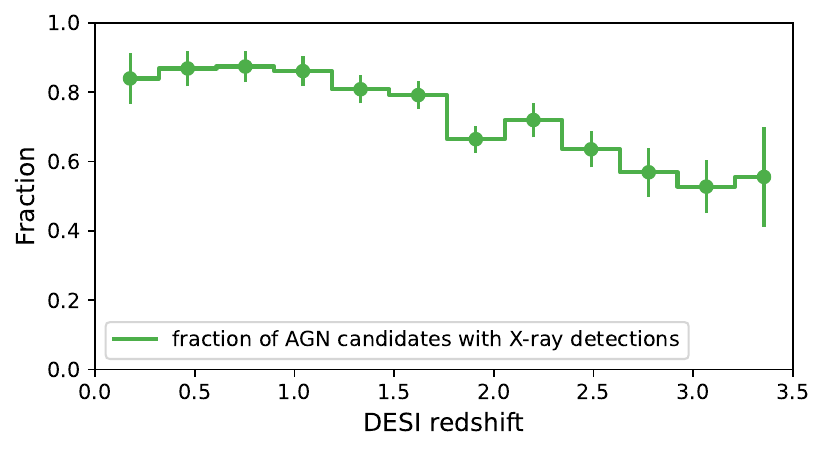}
    \includegraphics[width=0.99\linewidth]{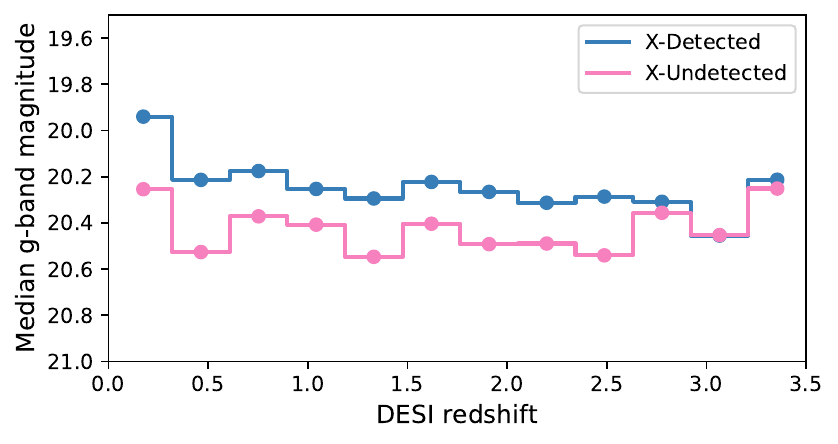}

    \caption{Redshift distribution of the variability-selected AGN candidates in the eFEDS field that have DESI spectra. Top: distributions of the X-Detected and X-Undetected samples, separated by their DESI pipeline classifications of {\tt QSO} or {\tt GALAXY}. Middle:  fraction of X-ray detections, classified as {\tt QSO} or {\tt GALAXY} by DESI, within the variability-selected AGN candidates. Bottom: Median ZTF g-band magnitude of X-Detected and X-Undetected DESI{\tt QSO} or {\tt GALAXY} sources.}
    \label{fig:redshifts}
\end{figure}

The redshift distributions of the \xd{} and \xu{} samples are significantly different, as can be seen in the top panel in Fig. \ref{fig:redshifts}, where the \xd{} sources cluster around $z\sim1.3$ and the \xu{} sources at $z\sim 2$. The middle panel in the figure shows the fraction of AGN candidates, classified as either {\tt QSO} or {\tt GALAXY} by DESI, that have X-ray detections. This fraction drops steadily from about 85\% at low redshift toward 50\% at $z\sim2-3$. This reduction in X-ray counterparts toward higher redshifts is unlikely to be simply a product of lower fluxes in otherwise equal AGN, since the median $g$-band magnitudes of both samples, shown in the bottom panel of the same figure, remain approximately constant with redshift, after the initial, lowest redshift bin. It is also clear that the \xu{} sample contains slightly dimmer objects than the \xd{} sample at the same redshift, by about 0.25 magnitudes in the $g$-band. This difference might explain a fraction of the variability selected AGN that fall below the eFEDS detection limit, but does not explain the redshift dependence. Finally, the small number of variability-selected AGN that are classified by DESI as {\tt GALAXY}, with or without X-ray counterparts, appear almost exclusively in the lowest redshift bins (i.e. $z<0.8$). These populations will be discussed further in Sec. \ref{sec:varAGN_galaxy}.

 The evolution of the X-ray-detected fraction points toward an evolution of the \emph{observed} optical to X-ray flux ratio, which might be an effect of the K-correction on an SED that rises towards higher energies in the optical/UV bands but drops towards higher energies in the X-ray band.  Alternatively, this result might show that the variability-selected AGN at higher redshifts, in an optical-flux-limited sample, are intrinsically X-ray dimmer. This is expected since more distant objects must be more luminous and the X-ray to UV flux ratio is known to drop toward higher luminosities \citep[e.g.][]{Lusso2016}. Interestingly, the evolution of the spectral energy distribution (SED) with luminosity differs for changes in luminosity due to changes in black hole mass or changes in accretion rate. Therefore, we use the SED fits from \citet{Chen25}, which are split for  a grid of black hole mass $M$ and Eddington ratio ($\lambda$ ), to gauge the evolution of X-ray detectability toward higher redshifts.
 
\begin{figure}
    \centering
    \includegraphics[width=0.99\linewidth]{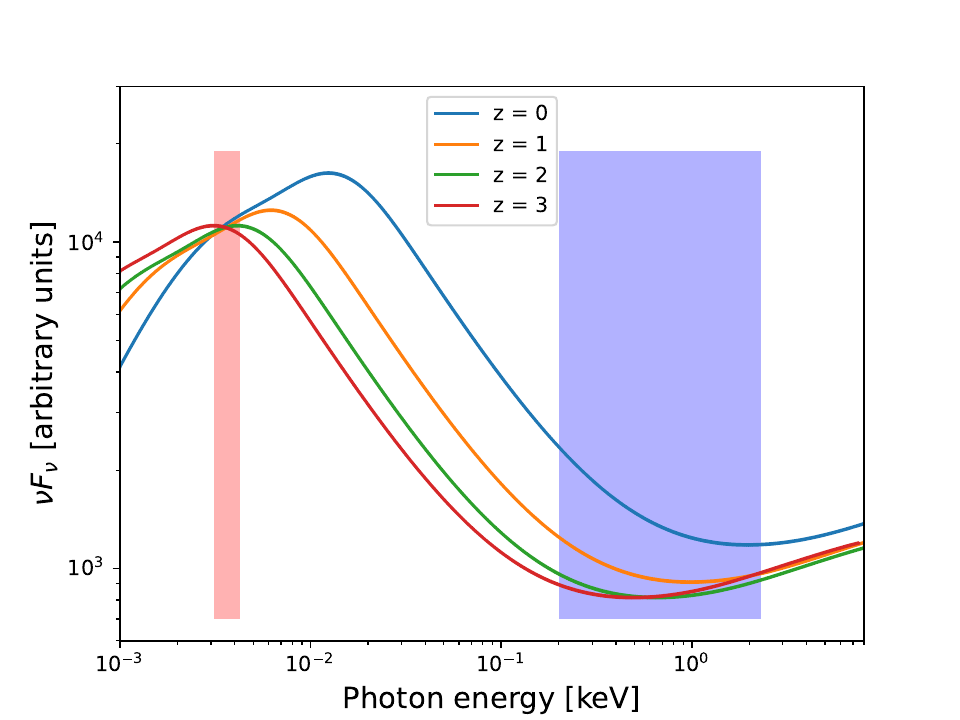}
    \caption{The strong X-ray K-correction: AGN spectral energy distribution modelled with the AGNSED model and parameters fit to the average fluxes measured for AGN with $log (M/M_\sun)=8.73$ and $\lambda=10\%$ in \citet{Chen25}, at four different redshifts and normalised at the g-band flux. The $g-$bandpass is marked in red and the 0.2--2.3 keV X-ray band is marked in blue. }
    \label{fig:model1}
\end{figure}

 We adopt for the moment the parameters from the most populated bin in the grid in \citet{Chen25}, [$log(M/M 
\odot)=8.73; \lambda=10\%$] and use the \texttt{AGNSED} spectral model \citep{Kubota18}. By shifting this baseline SED to higher redshifts while fixing the observed g-band flux, we measure the resulting variation in X-ray flux (see Fig. \ref{fig:model1}). For this static SED shape, the \emph{observed} X-ray-to-optical flux ratio drops by approximately 50\% as an AGN is redshifted from $z=0$ to $z=1$, with an additional 20\% decrease by $z=2$. Notably, this ratio remains largely stable between $z=2$ and $z=3$. This contrasts with the fraction of X-ray detections shown in middle panel in Fig. \ref{fig:redshifts}, which appear approximately constant between $z=0$ and $z=1$, and then decrease by up to $\sim$25-30\% at $z\sim3$, implying that the redshift evolution of this static SED does not explain well the missing fraction.

\begin{figure}
    \centering
    \includegraphics[width=0.99\linewidth]{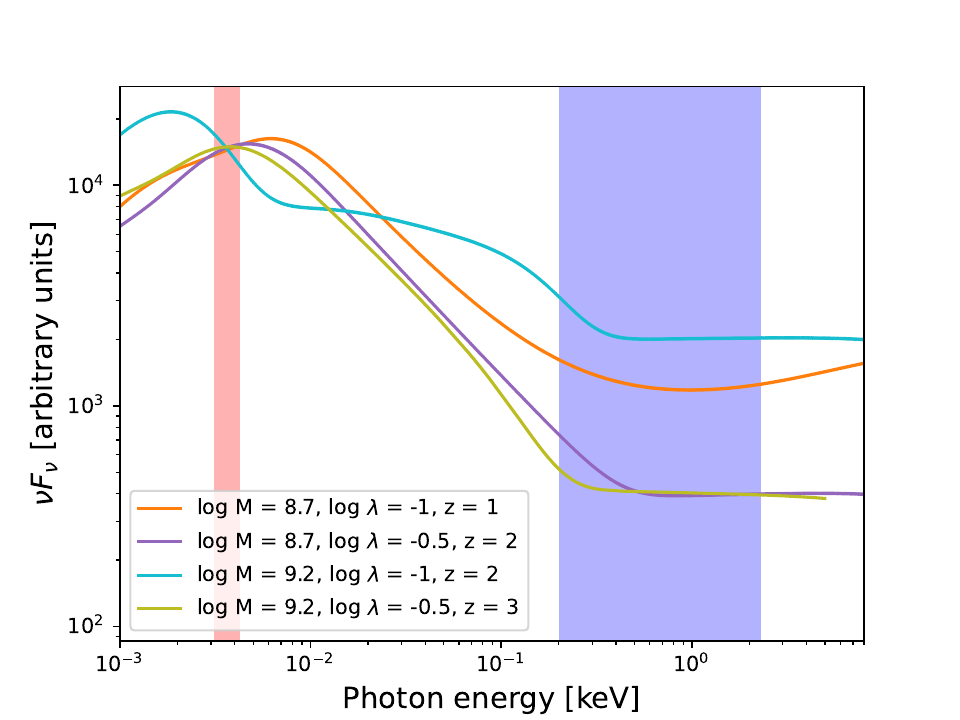}
    \caption{Mass- and Eddington ratio-dependence of the X-ray emitted fraction: AGN spectral energy distribution modelled with the AGNSED model and parameters fit by \citet{Chen25} for AGN with the average black hole masses and Eddington ratios noted in the legend. These have been redshifted to $z=2$ or $z=3$ and scaled to maintain equal fluxes in the observed $g$-band, which is marked by the red box. }
    \label{fig:model_many}
\end{figure}

However, because more distant AGN must be intrinsically more luminous to remain detectable, they are typically more massive and/or more rapidly accreting. Since the SED shape itself depends on these physical parameters, we evaluate the impact of an evolving SED in Fig. \ref{fig:model_many}. We compare our $z=1$ baseline to two $z=2$ scenarios: one with a threefold increase in $M$ and another with a threefold increase in $\lambda$. In all cases the SEDs are normalized to maintain the same observed $g$-band flux, to stress the effect of a varying SED shape on the X-ray detectability of optically-detected AGN. For comparison, we also include the model fit to the highest mass and highest Eddington ratio bin of \citet{Chen25}, and place it at $z=3$. These extreme objects have the same X-ray weakness for a given optical flux as the case with high Eddington ratio described above.  

The results indicate that SED evolution significantly affects their X-ray detectability at high redshift. At $z=2$, higher-mass/lower-Eddington ratio sources remain relatively accessible in the X-rays --- even slightly enhanced compared to the $z=1$ baseline --- whereas lower-mass/higher-Eddington ratio AGN become significantly X-ray fainter. Consequently, while differential K-corrections between the optical and X-ray bands account for a general decline in X-ray detections at high redshift, the intrinsic variation of the SED with mass and accretion rate predicts the specific dropout of high-Eddington ratio sources from the \xd\ sample over the same redshift span.

The redshift distributions of the \xd{} and \xu{} samples differ significantly. Since many AGN properties in a \emph{optical-flux-limited} sample are redshift-dependent, all subsequent analyses are performed within narrow redshift bins.

\subsection{Differences in Eddington ratio }
\label{sec:REdd}
Motivated by the discussion above, we compare estimates of the Eddington ratio $\lambda$ for the \xd{} and \xu{} samples. Specifically, we adopt single-epoch black hole mass estimates based on Mg\,{\sc ii} broad emission lines from the DESI BHMass v1.7 Value Added Catalogue~\footnote{\url{https://data.desi.lbl.gov/doc/releases/dr1/vac/qmassiron/}}, following the prescription of \citet{Pan2025}, which includes corrections for Fe emission. This catalogue used the same DESI spectra analysed throughout this work, although Mg\,{\sc ii}-based masses are given only in the $0.6<z<1.6$ redshift range, providing estimates for 1215 \xd{} objects and 220 \xu{}. 

\begin{figure}[h]
    \includegraphics[width=0.99\linewidth]{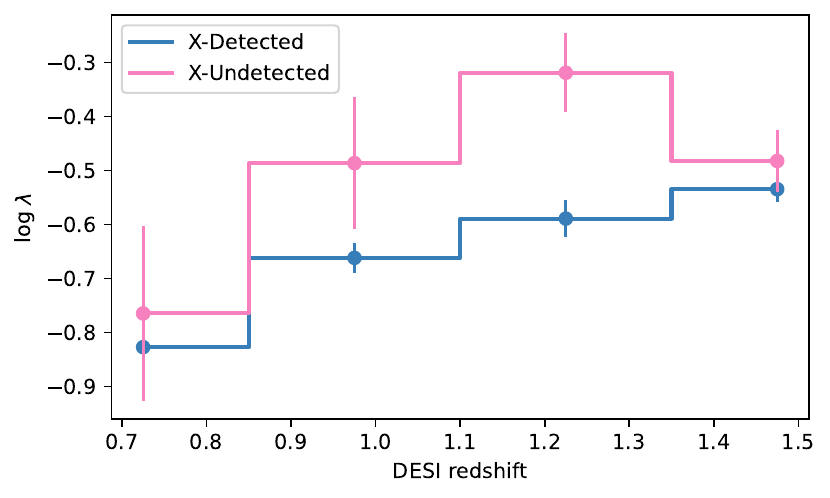}
    \includegraphics[width=0.99\linewidth]{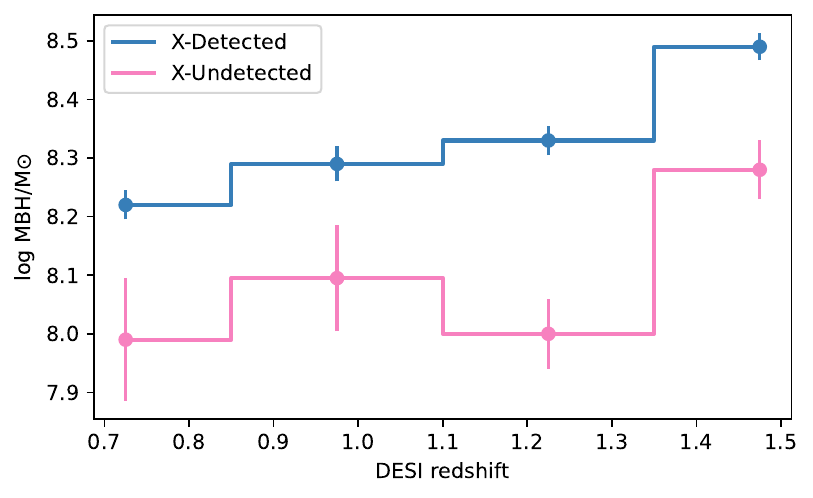}
    \caption{Top: Median values of the Eddington  ratio $\lambda$ of the  \xd{} and  \xu{} samples in four different redshift bins. Bottom: Median black hole masses for these samples in the same redshift bins.  
}
    \label{fig:REdd}
\end{figure}

We combined these masses with the luminosity at 3000\AA, published in the same catalogue, to compute Eddington ratios as $\lambda= BC\times L_{3000}/1.3\times10^{38}$M$_{\rm Pan}$, where the bolometric correction was $BC=3.7$ \citep[see table 1 in][for the $BC_{3000}$]{Chen2025} and M$_{\rm Pan}$ is in units of solar masses. 
In the top panel of Fig. \ref{fig:REdd} we show the median $\lambda$ of each sample, in four redshift bins. Although the error bars overlap on some bins, the Eddington ratios of the \xu{} sample are consistently higher, at least in this redshift range. For completeness we also show the median black hole masses in the same redshift bins in the bottom panel of Fig. \ref{fig:REdd}. These plots show that the \xu{} sample has median masses about a factor of 2 lower than the \xd{}, and correspondingly a factor of 2 higher Eddington ratios. 

\subsection{Differences in the optical spectra}
We searched for spectral differences in the \xu{} and \xd{} samples by building the median and several percentile spectra (i.e. the percentiles of the flux as a function of wavelength for all spectra that fall in a class). For this visualisation, we downloaded the best DESI DR1 spectra for all these sources from the SPectra Analysis and Retrievable Catalog Lab \citep[SPARCL;][]{Juneau2025} as part of the Astro Data Lab \citep{Fitzpatrick2014,Nikutta2020,Juneau2021}. 

The objects in both samples were split into the same redshift bins. For each sample and bin, the DESI spectra were normalized to unity at a reference wavelength falling close to the centre of the rest-frame wavelengths covered for each redshift bin and away from emission lines, and subsequently interpolated to a common grid of wavelengths. The spectra were slightly smoothed with a 2-pixel FWHM Gaussian convolution kernel, and the median and other percentile fluxes were obtained for each wavelength bin. The wavelength range was trimmed to ensure that all the wavelengths had contributions from all spectra in the redshift bin, by imposing the limits $\lambda _{min}=3600\,\AA/(1+z_{min})$ and $\lambda _{max}=9880\,\AA/(1+z_{max})$. 

\begin{figure*}[h]
    \centering
    \includegraphics[width=0.49\linewidth,trim={2cm 0 2.9cm 0},clip]{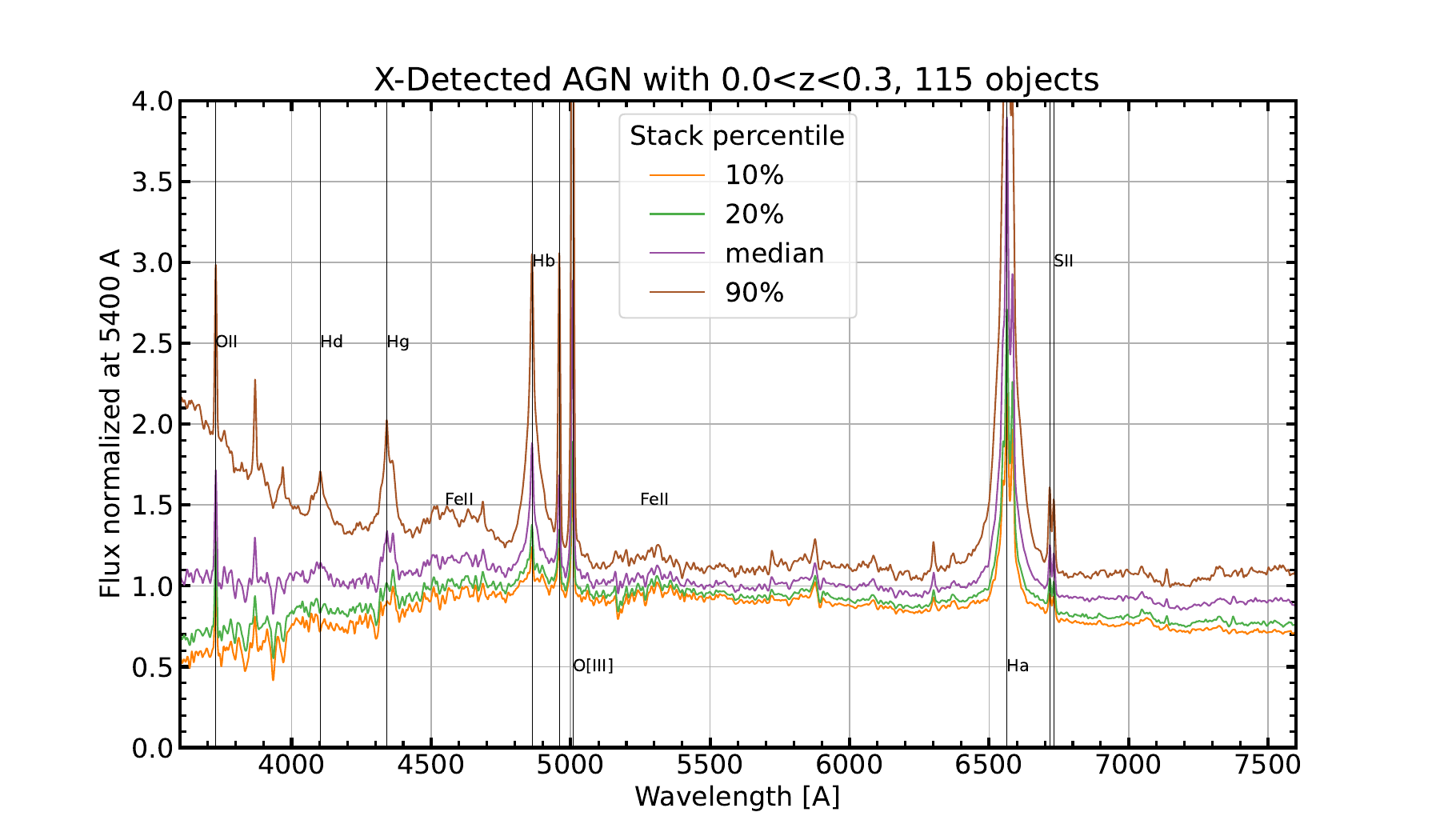}
    \includegraphics[width=0.49\linewidth,trim={2cm 0 2.9cm 0},clip]{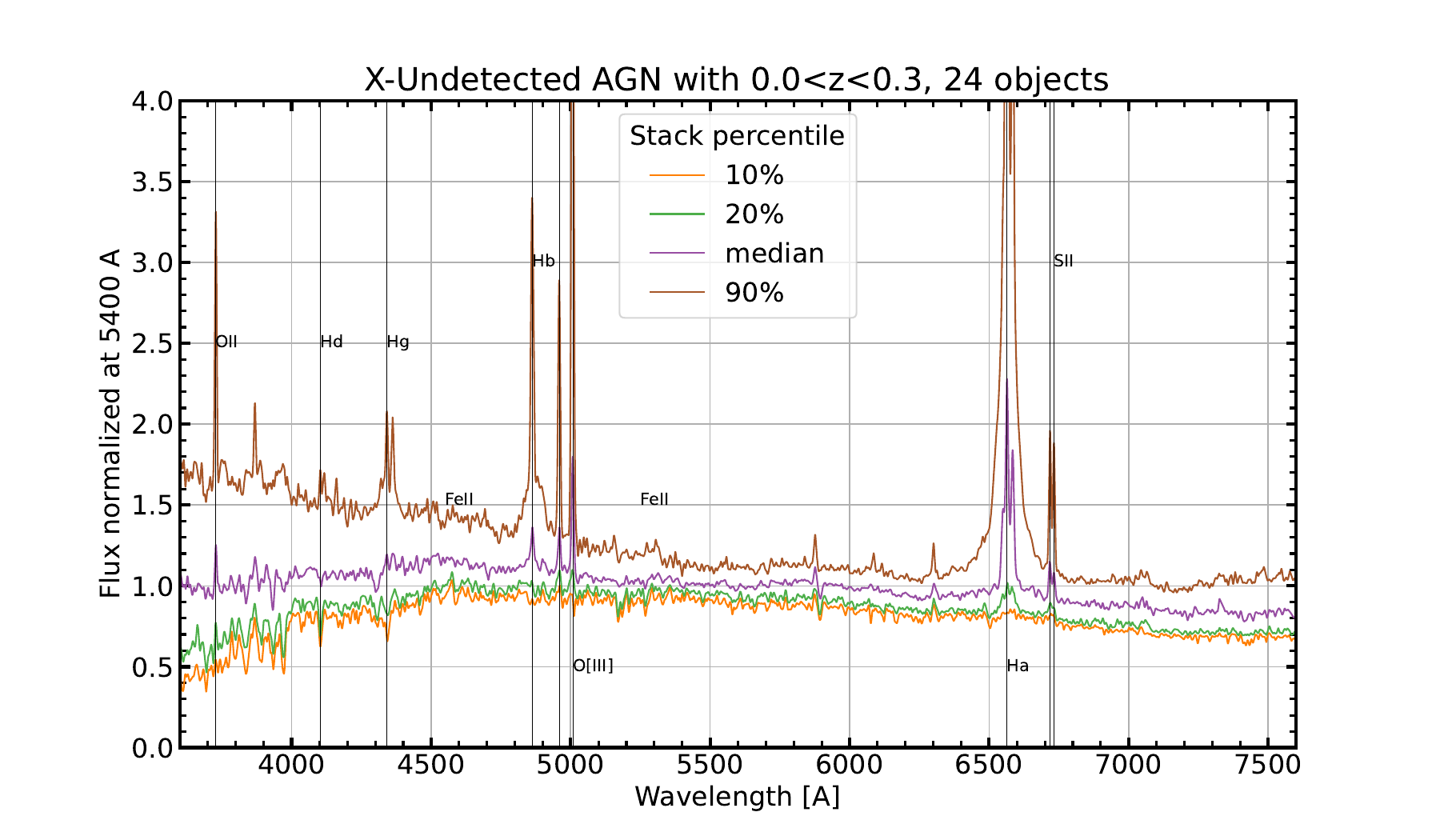}
    \caption{Median and percentile spectra for variability-selected AGN in the eFEDS field, with X-ray counterparts on the left and without counterparts on the right. The wavelengths are in the rest-frame, calculated using the redshift provided by the DESI pipeline. The redshift bins used to select spectra for each plot are written in each title, together with the number of spectra included in each bin. The reference wavelength for normalizing the spectra are written in the y-axis label. This figure shows the lowest redshift bin.}
    \label{fig:median_spectra_lowz}
\end{figure*}

\begin{figure*}[h]
    \centering
    \includegraphics[width=0.49\textwidth,trim={2cm 0 2.9cm 0},clip]{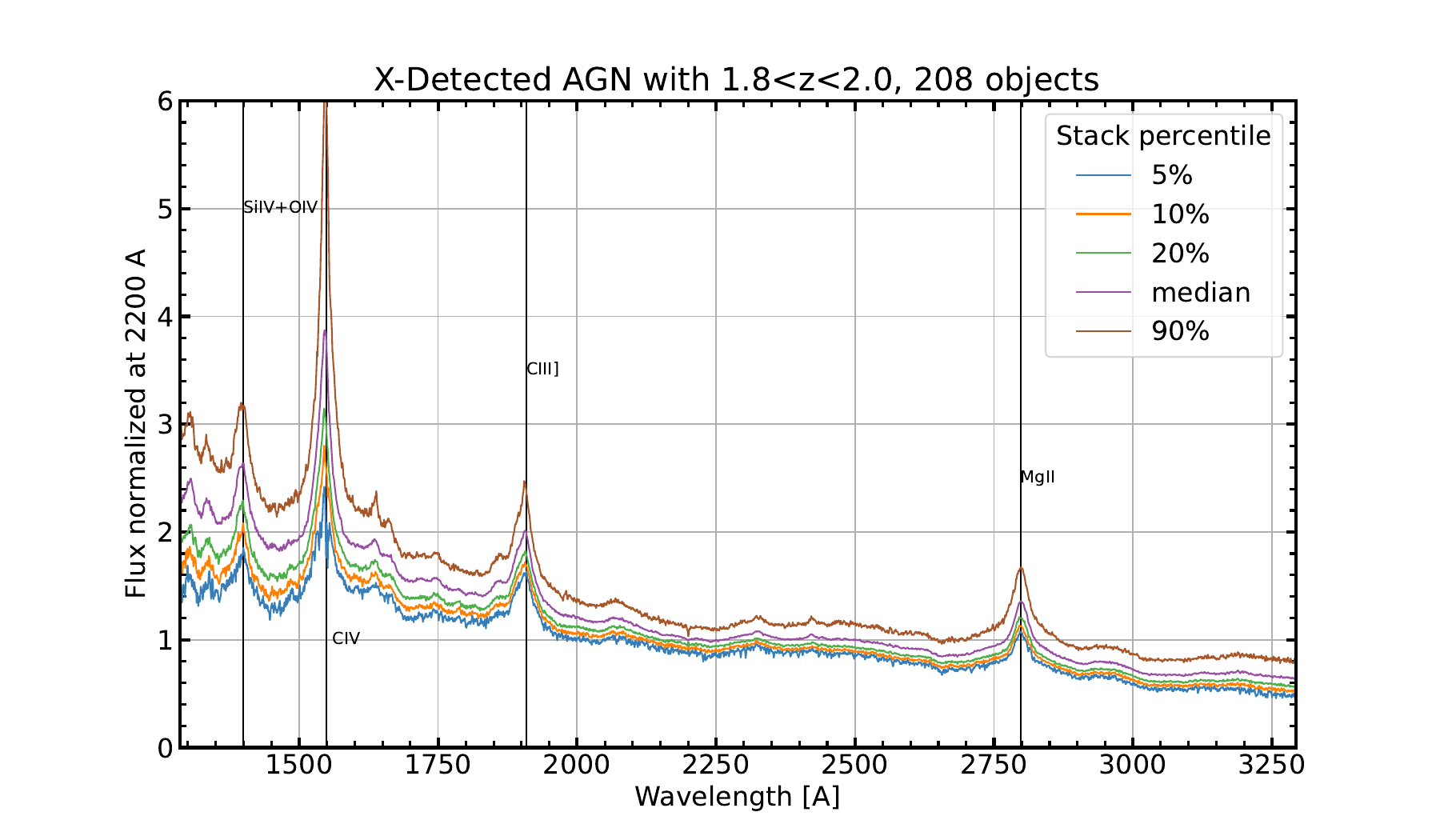}
    \includegraphics[width=0.49\textwidth,trim={2cm 0 2.9cm 0},clip]{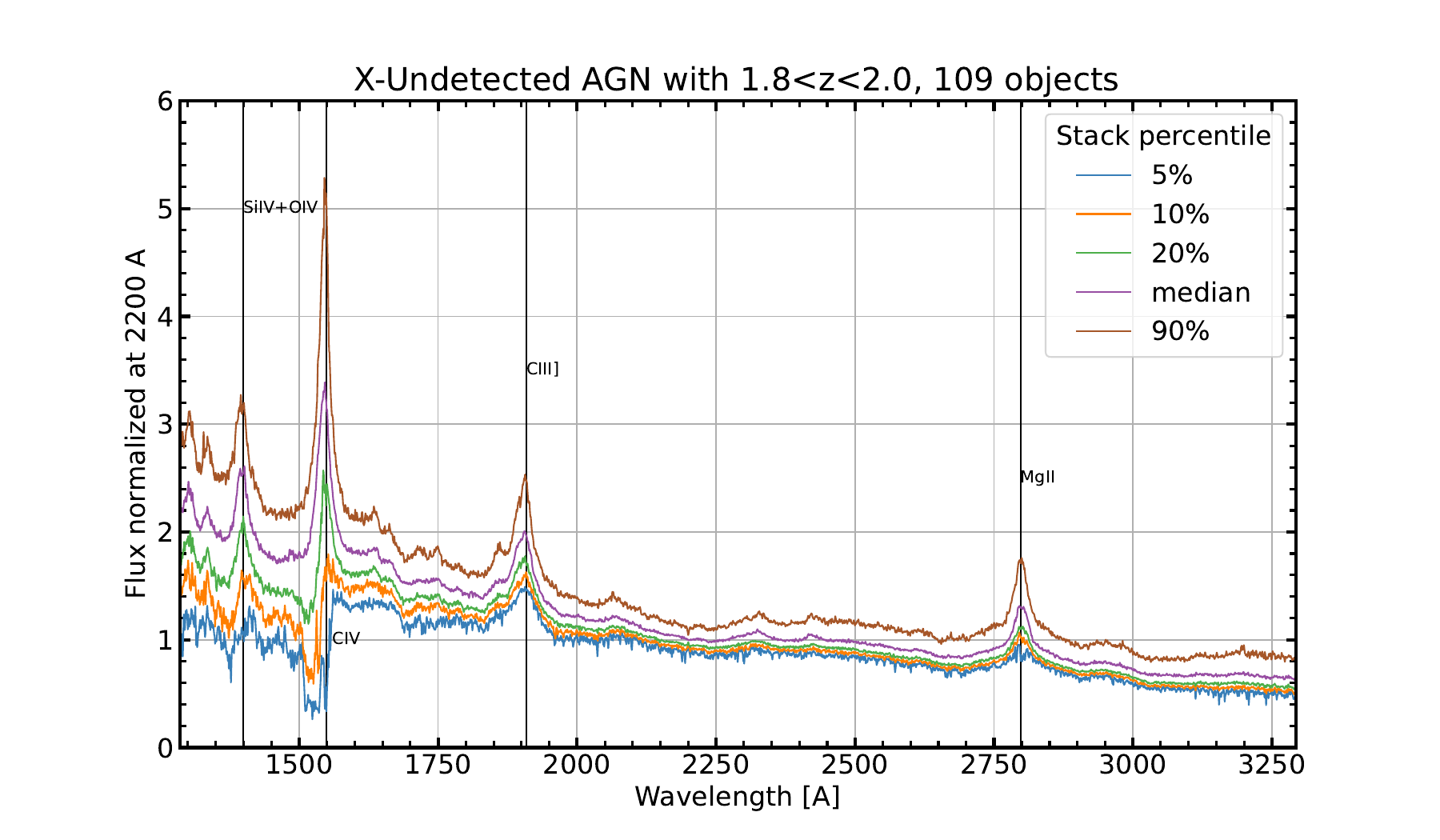}
    \caption{Same as Fig. \ref{fig:median_spectra_lowz} but for an intermediate redshift bin $1.8<z<2$, where the C\,{\sc iv} line is well sampled. }
    \label{fig:median_spectra_midz}
\end{figure*}

\begin{figure*}[h]
    \centering
    \includegraphics[width=0.49\textwidth,trim={2cm 0 2.9cm 0},clip]{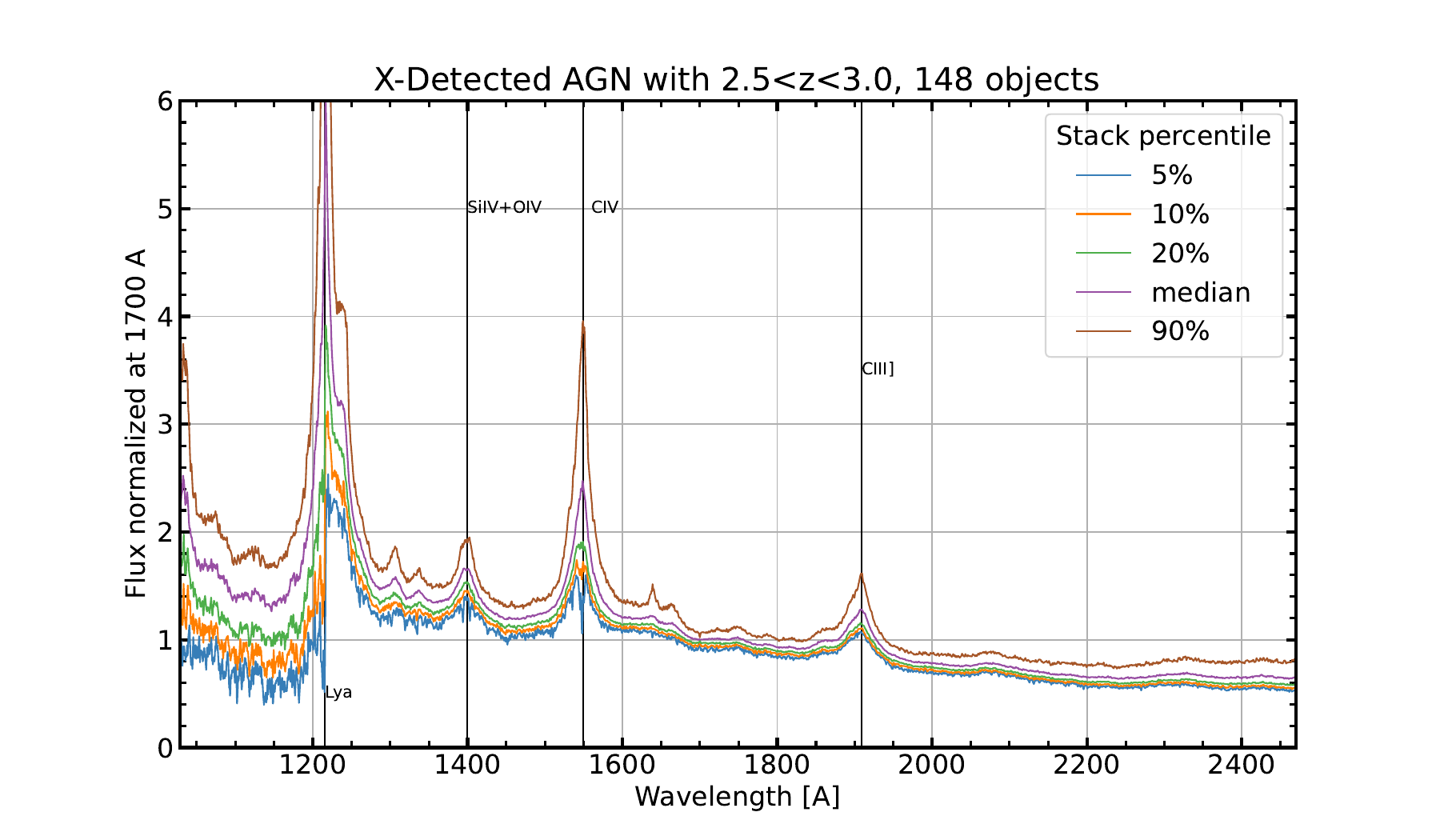}
    \includegraphics[width=0.49\textwidth,trim={2cm 0 2.9cm 0},clip]{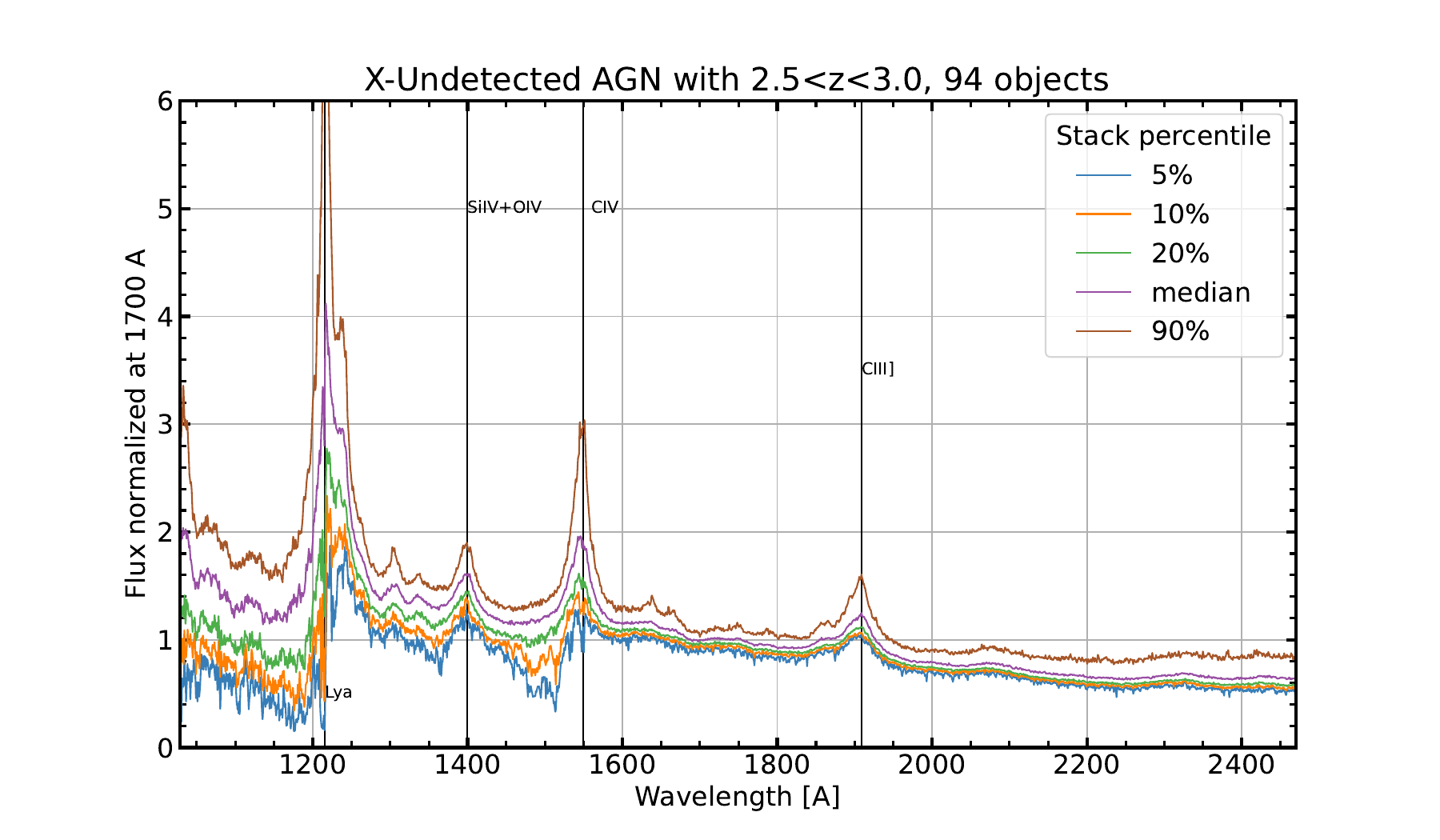}
    \caption{Same as Fig. \ref{fig:median_spectra_lowz} but for the redshift bin $2.5<z<3$. }
    \label{fig:median_spectra_highz}
\end{figure*}

The median spectra of different redshift bins, plotted in purple in Figs. \ref{fig:median_spectra_lowz}, \ref{fig:median_spectra_midz} and \ref{fig:median_spectra_highz} are  similar for the \xd{} (left column in all plots) and the \xu{} (right column in all plots) samples. These figures correspond to selected redshift bins that highlight different aspects of the comparison. There are, however, significant differences, namely: 
\begin{enumerate}
    \item The \xd{}  sample includes objects with bluer continua, especially evident in the lowest redshift bins and in their higher percentile spectra.
    \item The percentile spectra of the \xd{} sample shows stronger broad emission lines, although broad lines are also clearly found in the \xu{} sample everywhere except the lower 10 percentile of the lowest redshift bin ($0.0<z<0.3$).
    \item The broad absorption lines, blue-ward of the C\,{\sc iv} and Si\,{\sc iv} lines, are evident only in the \xu{} sample, in their 5, 10 and 20 percentile spectra.
\end{enumerate}   

 To quantify the difference in continuum flux and emission line properties, we use the automated spectral fitting provided by DESI, through the FastSpecFit routine. We downloaded these tables for the bright and dark programs, which contain most of the observed sources, and cross-matched them to the \xd{} and \xu{} samples, obtaining 2510 and 908 matches, respectively. For the tables below, we use all matches, regardless of their pipeline classifications. We note that the objects classified by the DESI Redrock pipeline as {\tt GALAXY}, that make up about half the objects in the \xu{} sample in the lowest redshift bin, are included in the ensemble spectral properties discussed below. 
 
 Tables \ref{tab:EWp} and \ref{tab:EWf} list the median equivalent widths of different emission lines for the \xu{} and \xd{} samples, in different redshift bins. In all cases, only bins with at least 20 objects are included, the errors on the medians are estimated by bootstrapping and the error on the ratio is estimated by error propagation. 

\begin{table}
\caption{Median EW of permitted emission lines} 
\label{tab:EWp}
\begin{tabular}{lllll}
\toprule
$z_{min}$ & $z_{max}$  & X-Detected & X-Undetected & Ratio \\
\midrule
\multicolumn{5}{c} {broad H$\alpha$}\\
0.1 & 0.3 & $97.28 \pm 8.19$ & $62.85 \pm 32.12$ & $1.55 \pm 0.80$ \\
0.3 & 0.5 & $146.92 \pm 7.31$ & $94.13 \pm 21.51$ & $1.56 \pm 0.37$ \\
\midrule 
\multicolumn{5}{c} {broad H$\beta$}\\
\midrule 
0.3 & 0.6 & $36.07 \pm 1.53$ & $27.51 \pm 3.34$ & $1.31 \pm 0.17$ \\
0.6 & 0.8 & $43.17 \pm 1.70$ & $38.62 \pm 4.19$ & $1.12 \pm 0.13$ \\
\midrule
\multicolumn{5}{c} {Mg\,{\sc ii} 2796}\\
\midrule 
0.5 & 0.9 & $14.86 \pm 0.41$ & $8.36 \pm 1.44$ & $1.78 \pm 0.31$ \\
0.9 & 1.3 & $13.55 \pm 0.40$ & $9.82 \pm 0.72$ & $1.38 \pm 0.11$ \\
1.3 & 1.7 & $12.86 \pm 0.35$ & $9.74 \pm 0.65$ & $1.32 \pm 0.10$ \\
1.7 & 2.1 & $11.59 \pm 0.40$ & $10.52 \pm 0.63$ & $1.10 \pm 0.08$ \\
2.1 & 2.5 & $10.37 \pm 0.57$ & $7.92 \pm 0.74$ & $1.31 \pm 0.14$ \\
\midrule
\multicolumn{5}{c} {C\,{\sc iv} 1549}\\
\midrule 
1.6 & 2.0 & $42.07 \pm 1.36$ & $30.03 \pm 1.62$ & $1.40 \pm 0.09$ \\
1.9 & 2.3 & $43.24 \pm 1.72$ & $34.31 \pm 1.30$ & $1.26 \pm 0.07$ \\
2.3 & 2.6 & $37.50 \pm 1.24$ & $29.99 \pm 2.06$ & $1.25 \pm 0.10$ \\
2.7 & 3.0 & $36.71 \pm 2.22$ & $28.73 \pm 1.56$ & $1.28 \pm 0.10$ \\
\midrule
\multicolumn{5}{c} {OI 1304}\\
\midrule
1.8 & 2.4 & $1.84 \pm 0.11$ & $1.58 \pm 0.21$ & $1.16 \pm 0.17$ \\
2.4 & 2.9 & $0.18 \pm 0.15$ & $0.28 \pm 0.16$ & $0.65 \pm 0.64$ \\
\midrule
\multicolumn{5}{c} {Si\,{\sc iv} 1396}\\
\midrule 
1.8 & 2.4 & $8.80 \pm 0.35$ & $8.42 \pm 0.45$ & $1.05 \pm 0.07$ \\
2.4 & 2.9 & $7.09 \pm 0.29$ & $7.27 \pm 0.53$ & $0.98 \pm 0.08$ \\
2.9 & 3.5 & $6.65 \pm 0.44$ & $6.51 \pm 0.90$ & $1.02 \pm 0.16$ \\
\midrule
\multicolumn{5}{c} {Ly$\alpha$}\\
\midrule 
2.2 & 2.6 & $66.66 \pm 2.54$ & $58.53 \pm 1.90$ & $1.14 \pm 0.06$ \\
2.6 & 3.1 & $64.28 \pm 3.14$ & $52.17 \pm 5.59$ & $1.23 \pm 0.15$ \\
3.1 & 3.5 & $58.12 \pm 5.66$ & $40.45 \pm 7.15$ & $1.44 \pm 0.29$ \\
\bottomrule
\end{tabular}
\end{table}

\begin{table}
\caption{Median EW of forbidden and semi-forbidden emission lines}
\label{tab:EWf}
\begin{tabular}{lllll}
\toprule
$z_{min}$ & $z_{max}$ & X-Detected & X-Undetected & Ratio \\
\midrule
\multicolumn{5}{c} {[N\,{\sc ii}] 6548}\\
\midrule 
0.1 & 0.3 & $2.80 \pm 0.27$ & $2.56 \pm 0.73$ & $1.09 \pm 0.33$ \\
0.3 & 0.5 & $3.23 \pm 0.25$ & $4.24 \pm 1.05$ & $0.76 \pm 0.20$ \\
\midrule
\multicolumn{5}{c} {[O\,{\sc iii}] 5007}\\
\midrule
0.3 & 0.6 & $15.48 \pm 0.89$ & $18.52 \pm 5.10$ & $0.84 \pm 0.24$ \\
0.6 & 0.8 & $14.47 \pm 0.66$ & $14.02 \pm 1.60$ & $1.03 \pm 0.13$ \\
\midrule
\multicolumn{5}{c} {Si\,{\sc iii}]	1892} \\
\midrule 
0.9 & 1.3 & $8.47 \pm 0.26$ & $8.15 \pm 0.39$ & $1.04 \pm 0.06$ \\
1.3 & 1.7 & $10.37 \pm 0.19$ & $11.20 \pm 0.44$ & $0.93 \pm 0.04$ \\
1.7 & 2.2 & $9.59 \pm 0.21$ & $10.28 \pm 0.31$ & $0.93 \pm 0.03$ \\
2.2 & 2.6 & $6.76 \pm 0.27$ & $7.72 \pm 0.46$ & $0.88 \pm 0.06$ \\
2.6 & 3.0 & $5.79 \pm 0.38$ & $6.79 \pm 0.68$ & $0.85 \pm 0.10$ \\
\midrule
\multicolumn{5}{c} {C\,{\sc iii}] 1908}\\
\midrule 
0.9 & 1.3 & $16.37 \pm 0.40$ & $14.83 \pm 1.04$ & $1.10 \pm 0.08$ \\
1.3 & 1.7 & $13.93 \pm 0.32$ & $12.75 \pm 0.67$ & $1.09 \pm 0.06$ \\
1.7 & 2.2 & $13.54 \pm 0.42$ & $13.11 \pm 0.57$ & $1.03 \pm 0.06$ \\
2.2 & 2.6 & $15.17 \pm 0.56$ & $14.60 \pm 1.11$ & $1.04 \pm 0.09$ \\
2.6 & 3.0 & $14.10 \pm 0.73$ & $13.31 \pm 0.99$ & $1.06 \pm 0.10$ \\
\bottomrule
\end{tabular}
\end{table}

Comparing the ratios between the median EW of lines in the \xd{} sample to the median EW of the \xu{} sample, we note a strong difference between the behaviour of permitted and forbidden lines. While the forbidden and semi-forbidden lines are equally prominent in the \xd{} and \xu{} samples, the broad, permitted lines are consistently stronger in the \xd{} sample. 

\begin{figure}[htbp]
   \includegraphics[width=0.99\linewidth]{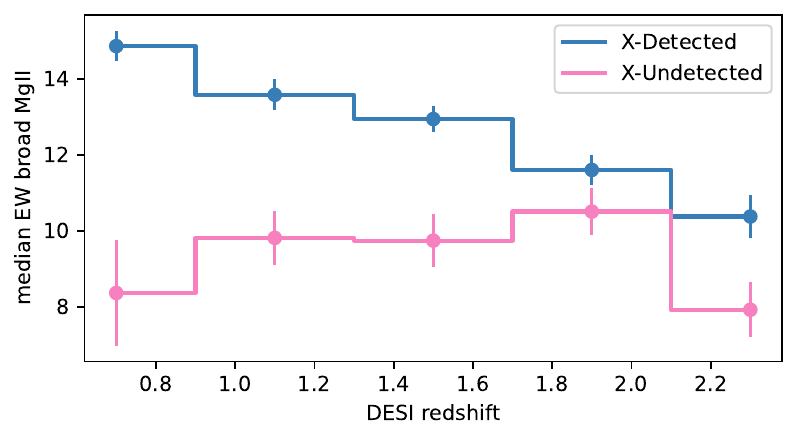}
   \includegraphics[width=0.99\linewidth]{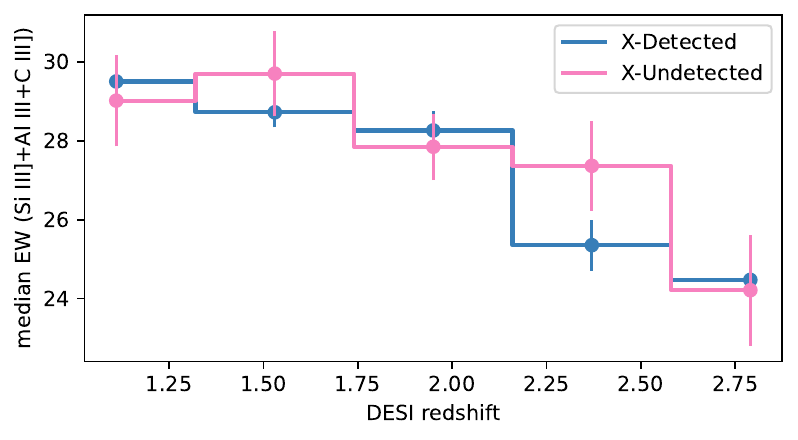}
    \caption{Median equivalent width of different emission lines, in the \xd{} (blue) and in the \xu{} (pink) samples, in different redshift bins.
    The top panel shows the results for the broad, permitted Mg II emission line and the bottom panel for the compound of semi-forbidden emission lines C\,{\sc iii}] and Si\,{\sc iii}], together with the permitted line Al\,{\sc iii}, which is blended with the former two but is subdominant. These lines are probed at similar redshift ranges.  }
    \label{fig:median_EW}
\end{figure}

In the lowest redshift bins, a comparison between the broad H$\alpha$ and H$\beta$ equivalent widths in both samples, shows that the \xd{} sample has significantly more prominent lines. For the same redshift bin (and therefore the same objects), the narrow emission lines of [N\,{\sc ii}] and [O\,{\sc iii}] show equal strengths in both samples, implying that the \xu{} sample might not simply be more contaminated by inactive galaxies, but that their broad line signatures are intrinsically weaker, and thereby more often classified as {\tt GALAXY} instead of {\tt QSO} by the DESI pipeline.  For all the other emission lines, which are detected from about $z>0.7$, the number of objects classified as {\tt GALAXY} by the DESI pipeline is negligible. For these lines too we note the same dichotomy: permitted, broad lines ($H\alpha$, H$\beta$, Mg\,{\sc ii} 2796, C\,{\sc iv} 1549 but not Si\,{\sc iv} 1396) are stronger in the \xd{} than the \xu{} samples, while forbidden and semi-forbidden lines for the same redshift bin have approximately equal strengths in both samples. To exemplify the trends of the emission line strengths, we plot in Fig.\ref{fig:median_EW} the cases for the permitted Mg\,{\sc ii}, broad  emission line (top) and the complex of semi-forbidden Si\,{\sc iii}] and C\,{\sc iii}] lines, together with the Al\,{\sc iii} permitted line, which is blended with the previous two but often much weaker (bottom). These plots have overlapping redshift ranges, for a more direct comparison of the difference between the behaviour of permitted and (semi-)forbidden lines. 
\subsubsection{Connection of emission line properties to differences in Eddington ratios}
The EW of C\,{\sc iv} is known to decrease with increasing Eddington ratio \citep[e.g.][]{Sulentic2007}, down to about 30 \AA , below which the Eddington ratio correlates more strongly with the velocity shift of the same line. The lower values of the EW in C\,{\sc iv} for the \xu{} sample therefore point to a higher median Eddington ratio in this sample. The values of the C\,{\sc iv} velocity shift obtained from FastSpecFit unfortunately reach the -2500 km s$^{-1}$ limit set for this parameter in the fitting routine, for many objects in both samples. However, we note that the C\,{\sc iv} velocity shift is slightly stronger in the \xu{} sample; when considering only objects with $1.6<z<3$, the median C\,{\sc iv} velocity shift is -2190 km/s for the \xu{} sample (463 objects) compared to -1970 km/s for the \xd{} sample (864 objects).

\begin{figure}[htbp]
   \includegraphics[width=0.99\linewidth,trim={2cm 0 2.4cm 0},clip]{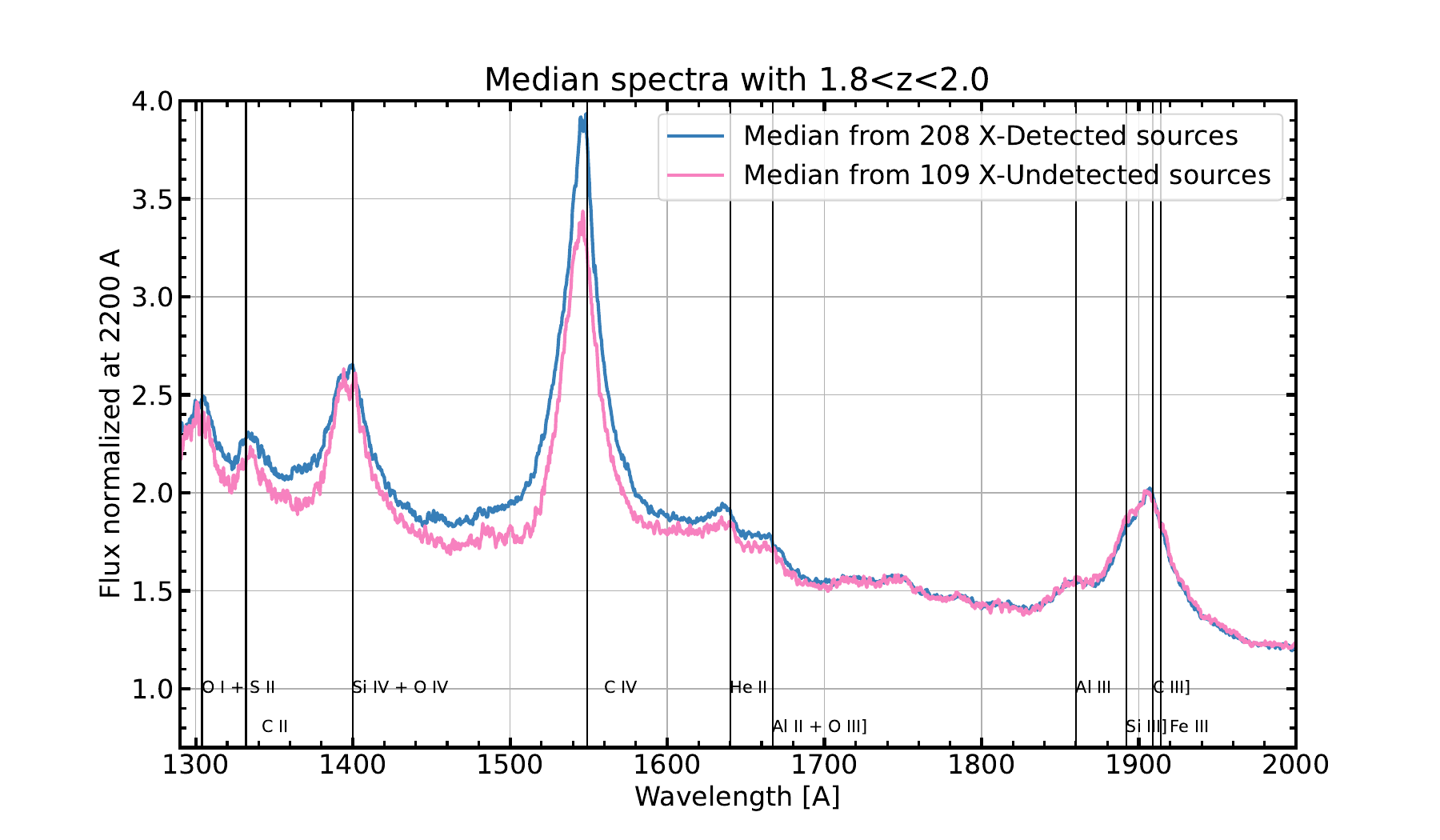}
    \caption{Median spectra of the \xd{} in blue and \xu{} in pink, for objects in the $1.8<z<2$ redshift range. }
    \label{fig:comparison_spectra}
\end{figure}

\citet{Wu2025} studied the evolution of the equivalent width of different broad lines with black hole mass and Eddington ratio for a sample of low redshift AGN. Their spectral fitting and photoionization model predictions both show that for masses above $10^8 M_{\odot}$, the EW of both C\,{\sc iv} and Mg\,{\sc ii} decrease with increasing Eddington ratio (see for example fig. 9 in their paper). This evolution could explain the difference we observe in the EW of the permitted emission lines, if the \xu{} sample indeed contained more higher accretion rate AGN, as discussed above and in Sec. \ref{sec:redshift}. Their calculations, however, predict the same behaviour for the C\,{\sc iii}] semi-forbidden line, which does not show a difference between the \xd{} and \xu{} samples, implying that the evolution of these lines with Eddington ratio, or the explanation for this dichotomy, may be more complex. A possible explanation could be related to the contributions of Al\,{\sc iii} and Fe\,{\sc iii}, which are blended with C\,{\sc iii}] and Si\,{\sc iii}],  making it difficult to reliably disentangle individual components \citep[e.g.][]{Martinez2018, Temple2020}. For example, if C\,{\sc iii}] is indeed weaker in the \xu{} objects, this could be compensated by a stronger Fe\,{\sc iii}, as expected for AGN with high Eddington ratios. Unfortunately, Fe\,{\sc iii} strengths are not reported in the FastSpecFit catalogue, so this hypothesis cannot be fully tested with current resources and is postponed to future work. Fig. \ref{fig:comparison_spectra} shows the location of these blended lines, together with a comparison of the median spectra of the \xd{} and \xu{} samples. This figure also highlights the similarity between both samples for most emission lines in this region except for C\,{\sc iv}.

The behaviour of the forbidden lines, however, is not consistent with the interpretation of higher accretion rates for the \xd{} sample. In previous studies, it has been shown that forbidden emission lines exhibit systematic variations across AGN with different accretion states. This behaviour has been extensively studied within the framework of the Eigenvector 1 sequence \citep{Boronson1992, Shen2014}. In this context, AGN with high accretion rates tend to display significantly weaker forbidden-line emission compared to those with lower accretion rates, whereas we find similar forbidden line EWs in both samples. 

\begin{figure}[htbp]
   \includegraphics[width=0.99\linewidth]{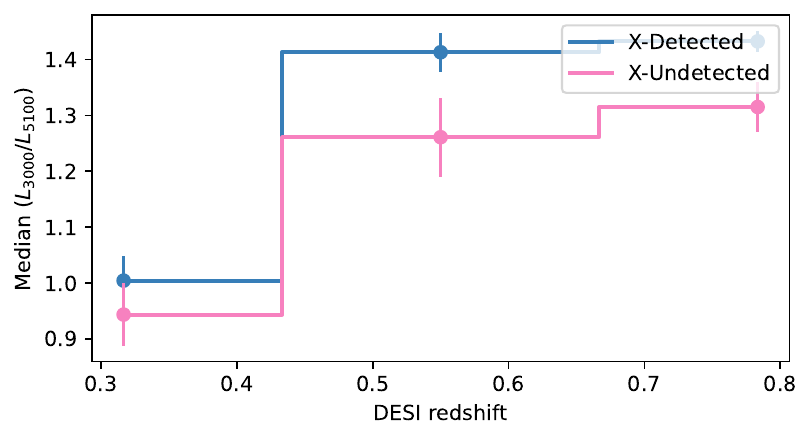}
   \includegraphics[width=0.99\linewidth]{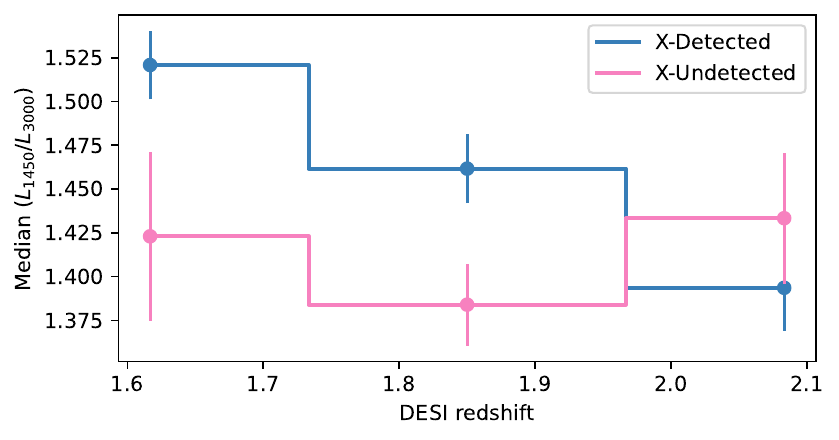}
    \caption{Median spectral slopes for the \xd{} and  \xu{} samples across different redshift bins, measured as the median of the ratios between the integrated luminosities at two different wavelengths, either 3000\AA/5100\AA\ (top) or 1450\AA/3000\AA\ (bottom).
    }
    \label{fig:median_optical_slope}
\end{figure}

The optical/UV spectra of the \xd{} sample also have slightly bluer continua, as measured by the median ratio between the integrated luminosities at 1400\AA\ and 3000\AA, for objects with $1.5<z<2.2$, and between 3000\AA\ and 5100\AA, for objects with $0.2<z<0.9$.  Figure~\ref{fig:median_optical_slope} shows the median of this ratio for each sample, across different redshift bins. Although the ratios show similar redshift evolutions, the \xd{} sample appears consistently bluer. 

We note that \citet{Pu2020} studied the fraction of X-ray weak AGN, restricting the analysis to non-BAL and non-radio loud objects. They find that among red QSOs  [i.e., those with $\Delta (g-i)>0.2$, where $\Delta (g-i)$ is the difference between the $g-i$ colour of an object and the mode of this colour for QSO at the same redshift], the X-ray-weak fraction is 13\%, which is much higher than the X-ray-weak fraction in QSOs with normal colours. The difference is even greater for weak-line QSOs (i.e., those with C\,{\sc iv} EW<16 \AA ) where the X-ray-weak fraction was found to be 35\%, in contrast with only 6\% in normal line QSO. These findings are in line with our results, where the \xu{} sample have slightly redder optical/UV continua and weaker broad emission lines, and suggests that these differences are not limited to the contributions of BALs to the \xu{} sample properties or radio-loud objects to the \xd{} sample properties. Their analysis was limited to QSOs in the $1.7<z<2.7$ range, without segregating their samples further by redshift, so it is not straightforward to compare the redshift evolution of the X-ray-weak sample studied here to their sample. 

\begin{figure}[htbp]
    \includegraphics[width=0.99\linewidth]{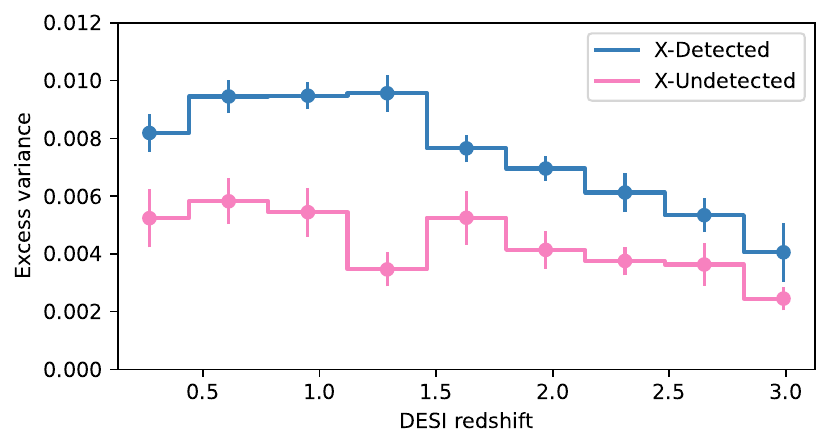}
    \caption{Total variability amplitude measured as the median excess variance for the \xd{} and \xu{} samples, across all studied redshift bins.}
    \label{fig:variability_properties}
\end{figure}

\subsection{BALs and X-ray flux}
We now focus on the stark difference in broad absorption line features, which are particularly apparent in the 5--20 percentile spectra of the \xu{} sample, shown in Figs. \ref{fig:median_spectra_midz}, \ref{fig:median_spectra_highz}, blue-ward of C\,{\sc iv}, Si\,{\sc iv} and Ly $\alpha$.
We note that any given spectrum participates in only one of these plots, so the different panels can be considered as independent experiments showing similar results.  

\begin{figure*}[h]
    \includegraphics[width=0.5\linewidth]{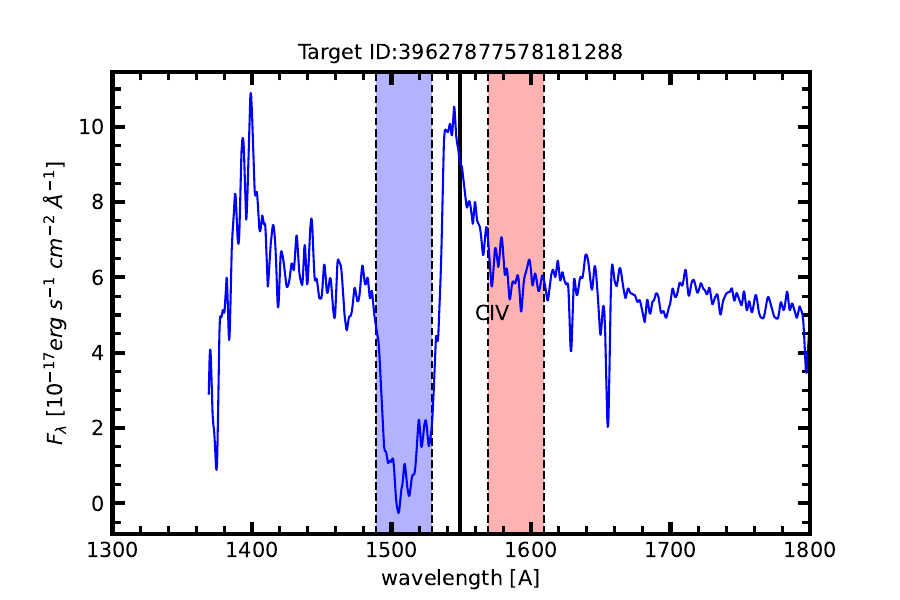}
    \includegraphics[width=0.5\linewidth]{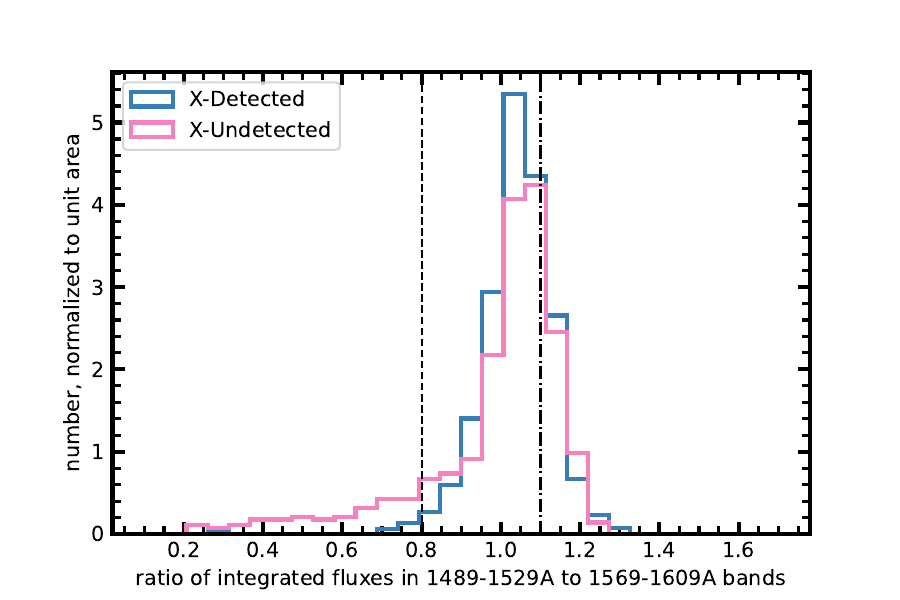}

    \caption{Left: example of the C\,{\sc iv} region of an \xu{} AGN showing broad absorption blue-ward of C\,{\sc iv}. The dashed lines mark the blue-ward and red-ward wavelength ranges used to integrate fluxes above and below the C\,{\sc iv} line. Right: distribution of the ratio of blue-ward over red-ward fluxes, for the \xd{} sample in pink and the \xu{} sample in blue. For both samples, only objects with $1.6<z<3.5$ were included. The dashed line marks the upper limit for sources considered as ``BAL'' for the subsequent X-ray stacking analysis while the dot-dashed line is the lower limit for the sources considered ``non-BAL''.  }
    \label{fig:BAL}
\end{figure*}

Figure \ref{fig:median_spectra_midz} shows the percentile spectra of objects in the $1.8<z<2$ redshift bin, \xd{} on the left and \xu{} on the right. The percentile spectra show that the largest difference between both samples appears around the 1500--1570 \AA\ wavelength range. In individual spectra, we note that the centre of the C\,{\sc iv} line is often blue-shifted, as seen for example in the left panel in Fig. \ref{fig:BAL}, and by different amounts for different objects, so simply taking the ratio between the flux towards higher and lower wavelengths of the nominal C\,{\sc iv} wavelength is not always a good indicator of absorption. Therefore, we defined wavelength windows as shown in Fig. \ref{fig:BAL} to integrate fluxes red-ward and blue-ward of the nominal centre of the C\,{\sc iv} line but discarding the region containing the line peak. We take the ratio between these fluxes, where low values likely correspond to strong absorption blue-ward of the line. The right panel in Fig. \ref{fig:BAL} shows the distribution of these flux ratios for objects with $1.6 <z< 3.5$.

We adopt the limits marked in the right panel of Fig. \ref{fig:BAL} to select 70 likely BAL (left of dash-dotted line) and 134 likely non-BAL AGN (right of dashed line) in the \xu{} sample. 
These limits are simply taken to select a sample in the tail to low values of the ratio (as compared to the \xd{} sample), and another sample at high values of the ratio, where the \xd{} and \xu{} samples have similar distributions.  

For the ``BAL'' and ``non-BAL'' samples, we extract eFEDS X-ray spectra at the optical positions. The extraction regions are circles of 0.6' radius for the source, and annuli between 1.2 and 2.4' for the backgrounds. Then, we stack the X-ray spectra of the two groups in the rest-frame with {\tt Xstack}\footnote{\url{https://github.com/AstroChensj/Xstack/}} \citep{Chen2025}, inputting also the spectroscopic redshifts and galactic column densities \citep[see][]{Liu2022}. {\tt Xstack} produces rest-frame observed stacked source and background spectra, and corresponding instrument responses that contain the galactic absorption. This enables X-ray spectral fitting with Poisson statistics and a z=0 model.

\begin{figure}[htbp]
    \includegraphics[width=5cm,height=0.9\linewidth,angle=-90]{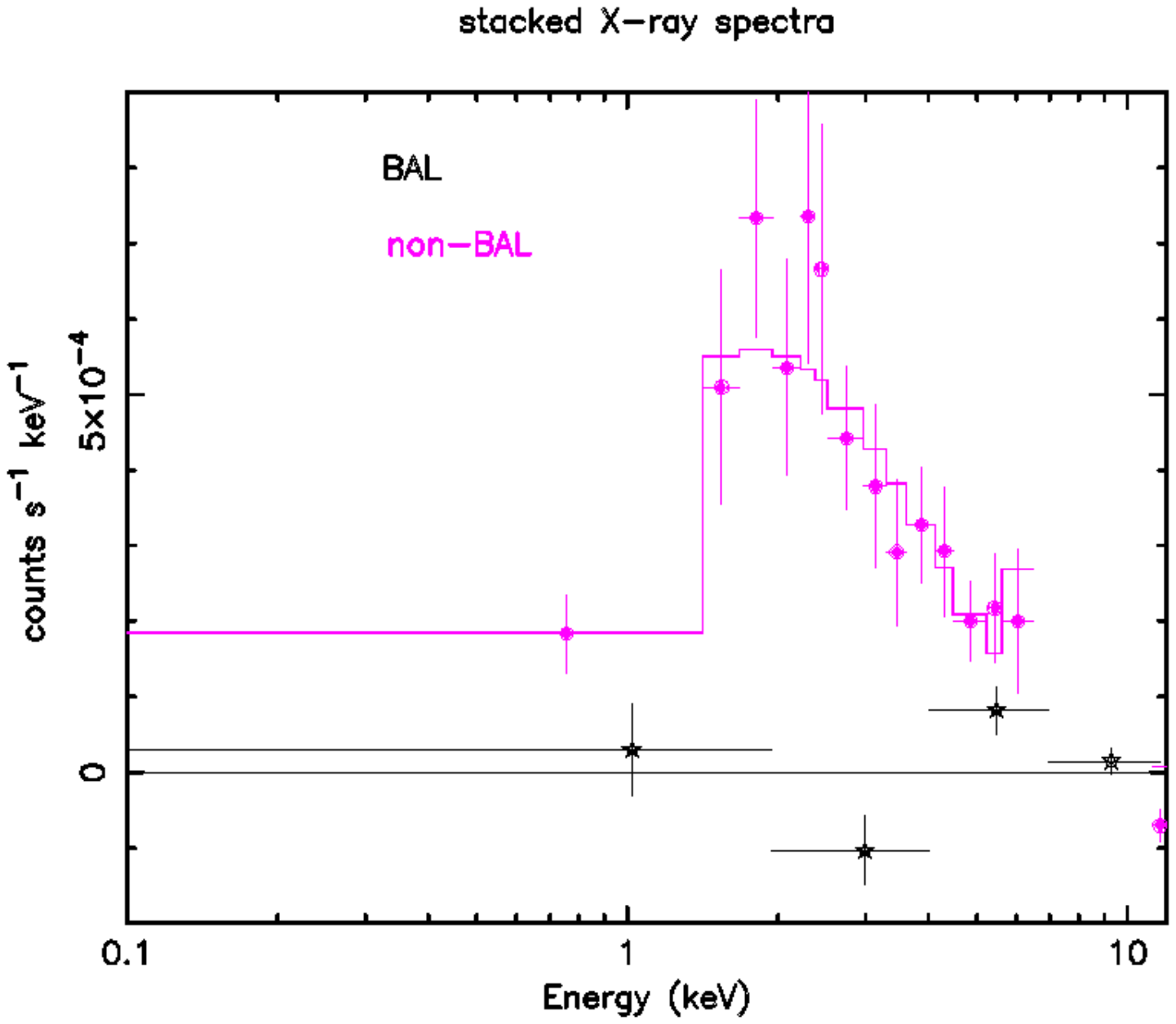}

    \caption{Restframe-stacked X-ray spectra from eFEDS for the \xu{} ``BAL'' sources, in black and ``non-BAL'' sources, in pink, both fitted with an absorbed powerlaw model with photon index $\Gamma=2$. The background-subtracted spectra are plotted in markers and best-fitting models in solid lines. }
    \label{fig:Xstack}
\end{figure}

The ``BAL'' group has 527 source region counts in the 0.5--5 keV \emph{rest-frame} energy range with 161 ks effective exposure, while the ``non-BAL'' group has 1477 source region counts with 301 ks effective exposure. The ``BAL'' group also has about half as many background counts, as expected from the scaled exposure. Comparing the source counts against the background, the ``BAL'' group, plotted in black in Fig. \ref{fig:Xstack}, appears consistent with the background, while the ``non-BAL'' group, plotted in pink, shows a significant detection of emission. Fitting with an absorbed powerlaw model ({\tt TBabs*powerlaw}), with a fixed Photon Index for the powerlaw of $\Gamma=1.9$ and tied absorption column densities, gives a normalization of $1.57\times 10^{-11}\pm1.8\times 10^{-12}$ for the ``non-BAL'' group while for the other spectrum the normalization is consistent with zero (i.e. $1.6\times 10^{-18}\pm3.2\times 10^{-12}$) and the absorbing column fitted value was $9\times 10^{20}\pm2\times 10^{21}$cm$^{-2}$ for both sets. The $1\sigma$ upper limit of the powerlaw normalization is therefore about 5 times lower than the measured normalization of the ``non-BAL'' set. We conclude from here that the \xu{}, ``non-BAL'' sample contains X-ray emitting AGN that happened to be just below the detection threshold of eFEDS, while the  \xu{}, ``BAL'' sample is significantly X-ray weaker.

We note that \citet{Brandt2000} found that absorption in the C\,{\sc iv} line correlates strongly with the optical to X-ray flux slope $\alpha_{OX}$ and that weak X-ray quasars have a higher probability of being BALs. Similarly, \citet{Blustin2008} report on the different $\alpha_{OX}$ of X-ray selected and optically-selected BALs, with X-ray selected BALs having significantly higher X-ray to optical fluxes, and \citet{Gibson2009} reports that BAL quasars are X-ray weaker than non-BAL counterparts, all in line with what we find. These results are not in conflict with the recent conclusion of \citet{Hiremath2025}, who find consistent X-ray properties in BAL and non-BAL samples, since their study is restricted to X-ray selected BALs, i.e. the subset that does have X-ray counterparts.

\subsection{Variability statistics}

Figure \ref{fig:variability_properties} shows the median excess variance obtained from the ZTF g-band forced photometry lightcurves, published in \citet{Arevalo2025}, as a function of redshift. For both samples, the variance drops towards higher redshifts, as expected in a flux-limited AGN sample due to several factors. For example, AGN samples at higher redshift are typically biased toward intrinsically more luminous sources. Considering the well-known anti-correlation between AGN luminosity and variability amplitude \citep[e.g.][]{Angione72, Hook94, Cristiani96,VandenBerk04}, this leads to lower observed variability in higher-redshift bins. In addition, the characteristic timescales of variability, increase both intrinsically with black hole mass and empirically with redshift (due to cosmological time dilation), making high redshift objects appear even less variable. More intriguing is the difference between the \xd{} and \xu{} samples within each redshift bin, where we note that the \xd{} sample is significantly more variable. Since the \xd{} are on average slightly more luminous (see bottom panel in Fig. \ref{fig:redshifts}), we would naively expect this class to be less variable, not more. Host-galaxy contamination can dilute the intrinsic variability, but this is not expected to be a factor at redshift above $z=1$, where the g band covers the 2000--2750\AA\ rest-frame wavelength range, making host contributions negligible. Other statistics that quantify the amplitude of the variability, such as the Damped Random Walk variance $\sigma_{DRW}$ \citep{Kelly09}, or the Mexican Hat power spectrum \citep{Arevalo12} on long timescales, show similar trends.  

Beyond its luminosity dependence, optical variability is well-known to anti-correlate with the Eddington ratio. Focusing on long-term variations (i.e., hundreds of days), both \citet{Sanchez-Saez18} and \citet{Arevalo2023} concluded that the amplitude of variability on long timescales is driven almost exclusively by the accretion rate rather than black hole mass. The observed decrease in variance within the \xu\ AGN sample is therefore consistent with our findings in Sec. \ref{sec:REdd} that high-Eddington ratio objects preferentially drop out of the X-ray catalogue.  Consequently, the difference in variability amplitude of the two samples can be explained by a difference in their median Eddington ratios, driven by these selection effects.

\subsection{Radio counterparts}
Radio emission is also a signature of accretion in some galaxies. In order to check whether X-ray detections are related to radio detections in the variability-selected AGN, we cross-matched the \xd{} and \xu{} samples to the Very Large Array Sky Survey (VLASS)
\citep{Lacy20} catalogue of reliably-detected sources in 2--4 GHz, presented by \citet{Gordon2021}. Cross-matching by optical coordinates, within a 1\farcs5 radius, showed that 
14/969=1.4\% objects in the \xu{} sample and 131/2886=4.5\% objects in the \xd{} sample have a radio counterpart. Therefore, although the fraction of radio-detected AGN are low in both samples, the \xd{} objects are 3 times more likely to have a radio counterpart in VLASS than the \xu{} objects, pointing to a higher average radio luminosity. This is qualitatively expected considering the good correlation between 5 GHz, nuclear radio luminosity, X-ray luminosity and black hole mass presented in \citet{Merloni2003}, since the \xd{} sample must have higher X-ray luminosities and possibly also higher black hole masses. 

\subsection{Effect of X-ray variability on the detection fraction}
As noted above, some variability-selected AGN did not have an X-ray counterpart in the eFEDS survey but were subsequently detected in the single scans of eROSITA, namely eRASS1 through eRASS4, which were $\approx$ 8 times shallower than eFEDS. Therefore, it is likely that some AGN missed an X-ray detection in eFEDS due to their intrinsic variability. Of the 1127 AGN candidates that did not match an eFEDS counterpart, 53, 45, 41, 27, matched a counterpart in eRASS 1, 2, 3 and 4, respectively, with some overlap objects detected in different scans. The total number of AGN candidates without eFEDS counterparts but with eRASS counterparts is 141, i.e. 12.5\% of the undetected sample. Only one of these was detected in all four scans, another one was detected in three scans, 20 were detected in two scans and the remaining 119 were detected only once. Of these 141 objects, 139 appear in the AGN/Galaxy Classification VAC and all of them are classified as broad line AGN.

\section{Complementary populations in the eFEDS field and classification discrepancies.}
 \label{Sec:complementary}
 Within the eFEDS area we have compared optical-variability AGN selection with a magnitude limit of $g\sim20.8$ to X-ray selected AGN with a eROSITA 0.2-2.3\,keV flux limit of $10^{-15}$\,erg/s/cm$^2$. This resulted in 2745 and 1127 variability-selected AGN with and without a matching eFEDS AGN counterpart, respectively, which have been discussed above, and about 10800 X-ray AGN with optical counterpart fluxes below the optical threshold of the ZTF lightcurves used for the selection of AGN candidates. These optically-dim, X-ray selected AGN could are beyond the scope of the present paper, which is based on magnitude-limited, optical-variability classifications. Below, we consider AGN which might be hidden amongst the optically bright galaxy population.

\subsection{Variability-selected AGN spectroscopically classified as {\tt GALAXY}}\label{sec:varAGN_galaxy}

As noted in Sec. \ref{section:results} and in the top panel in Fig. \ref{fig:redshifts}, at low redshifts ($z<0.6$), a significant fraction of the \xu{} sample are spectroscopically classified as {\tt GALAXY} by the DESI Redrock pipeline. These might correspond to objects with either spurious variability or weak AGN signatures in the optical spectrum that are not picked up by the pipeline routine. We therefore look for the classifications of all the variability-selected AGN in the AGN/Galaxy Classification VAC of DESI (Juneau et al. in preparation), which investigates in detail at least 10 well-known AGN spectral and photometric diagnostics to improve AGN classifications, particularly focusing on finding obscured, narrow-line, and weaker AGN populations. These fits were done for objects originally classified as {\tt QSO} or {\tt GALAXY}, so it does not give additional information on the ratio of stars misclassified as variable AGN. 

Removing the restriction of {\tt DELTACHI2} > 25, imposed above for the Redrock pipeline fits, increases slightly the number of objects. In the AGN/Galaxy Classification VAC, there are 807 entries for unique objects in the \xu{} sample, 42 of which were classified as {\tt GALAXY} by Redrock pipeline. All of them are at low redshift, where they represent about half of the \xu{} sample. Of these 42, 22 are re-classified as broad line AGN (i.e. with FWHM\,$>$\,1200\,km/s), 13 do not exhibit broad emission lines but are classified as AGN in at least one narrow-line diagnostic such as 
BPT \citep{Baldwin1981}, 
WHAN \citep{Cid2011} or 
BLUE  \citep{Lamareille2004,Lamareille2010} diagrams (i.e. non-broadline AGN), and 7 remain without AGN classification. This is consistent with the spectra in the right panel in Fig.~\ref{fig:median_spectra_lowz}, where the 10-percentile spectra of the \xu{} sample at $0<z<0.3$ shows no emission lines, but a small broad H$_\alpha$ component begins to be visible in the 20-percentile spectrum.

Similarly, in the \xd{} sample we find 67 objects originally classified by the Redrock pipeline as {\tt GALAXY}. Of these 67, 54 are reclassified as broad emission line AGN, 9 are AGN according to their narrow line characteristics and only 4 do not show AGN-like spectral features.  

We note that the limit of FWHM\,$>$\,1200\,km/s to determine the detection of broad lines can leave out some lower mass type I AGN, both in the \xd\ and \xu\ samples. Therefore, the non-broad line AGN selected through optical variability can correspond to either low mass AGN or to true type II AGN, where there is an unobscured view of the nucleus but the broad line region is absent. Of the 22 non-broadline AGN described above, six (three in each sample) have detected broad H$\beta$ lines, but their widths of about 300 km/s are too narrow to classify them as broad line AGN. These could correspond to type I AGN with low mass black holes.

\subsection{X-ray detected, non-variable galaxy}

Of the 3824 eFEDS AGN that have an optical variability classification, 766 are classified as non-variable galaxies. As already discussed in \citet{Arevalo2025}, these have about 2 dex lower X-ray luminosities than the eFEDS AGN that were also classified as AGN by their optical variability, showing that high X-ray luminosities also prompt higher levels of optical AGN signatures. Of these 766 objects, 572 appear in the DESI zall-pix-iron.fits catalog, almost exclusively with redshifts below 0.5 and a $z\sim0.3$ distribution peak, with 452/572 (79\%) classified as {\tt GALAXY}, 118/572 (20.1\%) as {\tt QSO}, and 2/572 as {\tt STAR} according to the Redrock pipeline.

In order to understand whether the objects classified as galaxies by their optical spectra correspond to weak type I AGN, type II AGN or star-forming or passive galaxies, we cross-matched the sample of 766 eFEDS AGN classified as non-variable galaxy, to the AGN/Galaxy Classification VAC. We found 570 matches with good spectra, 160 of which already had a {\tt QSO} classification from the DESI Redrock pipeline and/or had broad emission lines (FWHM\,$\ge$\,1200\,km/s) identified in this new catalog (28\%); 229 are classified as non-broadline AGN in this new catalog (40\%), and 181 objects had no AGN classification, and therefore are classified as star-forming or passive galaxies (32\%). 

This category, i.e. X-ray detections matched with non-variable galaxies, is where Type II AGN are expected to reside. In these sources, hard X-rays can penetrate the obscuration, whereas the variable optical emission from the accretion disc and the broad-line region remain hidden. However, we find that only 40\% of these mismatches actually correspond to non-broad-line AGN. The remaining cases are equally distributed between two types of misclassifications: Type I AGN erroneously classified as non-variable galaxies, and non-active galaxies incorrectly identified as X-ray AGN. Given that the total populations of true active galaxies identified as X-ray counterparts, and true non-active galaxies identified as non-variable galaxies, are significantly larger, these misclassification rates remain negligible. Nevertheless, when specifically using the mismatch between optical variability and X-ray detections to search for Type II AGN, this 40\% success rate must be taken into account. These few type I AGN that do not show optical variability and the non-active galaxies that show X-ray emission will be studied further in a separate work.  

\subsection{X-ray undetected, non-variable galaxy}

For completeness, we note that there are about 58,000 sources in the eFEDS area that \citet{Arevalo2025} have classified as non-variable galaxy and that do not appear in the eFEDS AGN catalog. Good quality spectra, selected using the same quality criteria as above, are available from DESI DR1 for 42,498 of these objects. Their DESI Redrock pipeline classification corresponds to {\tt GALAXY} for 42,100 (99\%) objects, {\tt QSO} for 155 (0.4\%) and {\tt STAR} for 243 (0.6\%). Alternatively, there are 43,088 of these objects in the AGN/Galaxy Classification VAC, with classifications of broad line AGN for 178 (0.4\%), non-broad line AGN for 10,434 (24.2\%) according to optical and/or UV emission line diagnostics, and the remaining 32,476 objects (75.4\%) are classified as galaxy with no optical/UV AGN features. This analysis shows that the objects that (i) are bright enough to have an optical variability-based classification, (ii) are classified as non-variable galaxy and (iii) have no X-ray counterpart, have a very small probability ($<1\%$) of having type 1 AGN optical spectra. 

In Table \ref{tab:summary} we summarize the confirmation fractions of these different populations. We separate the DESI Redrock pipeline classification of {\tt QSO} or {\tt GALAXY} from the AGN/Galaxy Classification VAC results, including broad line and non-broad line AGN. We note that the objects included in both catalogues differ slightly because in the first case the {\tt DELTACHI2} > 25 restriction was applied and in the second, only objects  classified as {\tt QSO} or {\tt GALAXY} were analysed.

\begin{table*}
    \caption{Summary of classifications}
    \centering
    \begin{tabular}{cc|l}
      Optical  & X-ray &  DESI spectral classification\\
      \hline
        AGN & detected &  98\% {\tt QSO}, 2\% {\tt GALAXY}, of 2,744 objects. \\
        && 99.6\% broad line QSO, 0.3\% non-broad line AGN, 0.15\% SF/passive galaxy, of 2,747 objects. \\
        \hline
        AGN & not detected & 92\% {\tt QSO}, 5\% {\tt GALAXY}, of 830 objects.\\
        && 97\% broad line QSO, 1.6\% non-broad line AGN, 1.1\% SF/passive galaxy, of 807 objects.\\
        \hline
        non-var galaxy & detected & 20\% {\tt QSO}, 79\% {\tt GALAXY}, of 572 objects.\\
        &&28\% broad line QSO, 40\% non-broad line AGN, 32\% SF/passive galaxy, of 570 objects.\\
        \hline
        non-var galaxy & not detected & 0.4\% {\tt QSO}, 99\% {\tt GALAXY}, of 42,498 objects. \\
        && 0.4\% broad line QSO, 24\% non-broad line AGN, 75\% SF/passive galaxy, of  43,088 objects.\\
    \end{tabular}
    \label{tab:summary}
\end{table*}

\section{Conclusion}
In this work, we analyze the properties of variability-selected AGN candidates in the eFEDS field, as presented by \citet{Arevalo2025}. We characterized these objects using public optical spectra from DESI DR1, which are available for 3,585 of the 3,872 variability-selected AGN candidates in this region. The classification purity is high: 3,523 sources are confirmed as broad-line AGN/QSO, 22 as non-broad-line AGN, and the remaining 40 are identified as "other" (either stars or non-active galaxies). 

Furthermore, many sources in this region were classified via variability as non-variable galaxies. The vast majority of these (approximately 99\% of those with DESI spectra) carry a DESI Redrock pipeline classification of {\tt GALAXY} while only $\sim 0.4\%$ are classified as {\tt QSO}.

The eFEDS field provides deep, uniform X-ray coverage from the eROSITA telescope. X-ray data were used to independently select AGN in this region, their optical counterparts were identified by \citet{Salvato2022} and their X-ray properties were analysed by \citet{Liu2022}. The overlap with the variability selection is mainly limited by the depth of the optical ZTF data. Of the $22,079$ X-ray AGN, optical variability only classified $3,824$ into any class, meaning most X-ray sources fell below the required optical flux threshold. Notably, despite the optical faintness of most of the X-ray AGN, approximately one-quarter of the variability-selected, optical AGN lack an X-ray counterpart in eFEDS. We also searched for X-ray detections with eROSITA during its four scans over the eFEDS field performed later as part of the All Sky Survey. In total, 141 variability-selected AGN that did not have a counterpart in eFEDS were subsequently detected in at least one scan. This is noteworthy since each individual scan is about 8 times shallower than eFEDS, showing that intrinsic variability of the AGN can prevent the detection of some objects in any given survey. These 141 objects were incorporated into the X-ray detected sample.  

As summarized in Table \ref{tab:summary}, our comparison between optical variability and X-ray selection reveals four distinct populations: 
\begin{enumerate}
    \item Dual-selected AGN: Sources identified by both methods have a 99\% probability of being classified as broad-line AGN.
    \item X-ray AGN / Optically Non-variable: This group contains a mix of 40\% non-broad-line AGN (the expected population for this mismatch), 28\% broad-line AGN missed by variability analysis, and 30\% that appear to be non-active by their optical spectrum.
    \item Optically variable AGN / X-ray Undetected: These are primarily broad-line AGN (95\%), with a small fraction of non-broad-line AGN (1.3\%) and a remainder of stars or non-active galaxies.
    \item Field Galaxies: The majority of galaxies in the field are classified as non-variable and remain undetected in X-rays. Over 99\% are spectrally confirmed as galaxies ---of which about a quarter correspond to non-broad line AGN, LINER or composite--- with the remainder 1\% split between stars and QSOs.
\end{enumerate}
The total number of objects with spectra in each category can be found in Table \ref{tab:summary}, while the total numbers, regardless of the availability of spectra are described in Sec. \ref{Sec:complementary}.

We conducted an in-depth analysis of the 3,782 variability-selected AGN, comparing those detected in X-rays ($N=2,886$) to those that are not ($N=969$). Both samples show a very high probability of having AGN-like spectra, though the probability is slightly higher for the \xd{} sample (see Table \ref{tab:summary}).

The redshift distributions of the two samples differ significantly: the \xd{} sample peaks at $z \approx 1$, whereas the \xu{} sample peaks at $z \approx 2.0$. Notably, about half of the variability-selected AGN at $z = 2.5-3.5$ lack X-ray detections. This "missing fraction" must be accounted for when extrapolating X-ray-selected populations to Type I AGN at high redshifts. We find that this redshift evolution can be explained by two factors: the varying K-corrections between optical and X-ray bands, and the evolution of the SED with luminosity. Both effects make X-ray detection increasingly difficult at high redshift, for a given optical flux. 

Interestingly, the impact of luminosity on X-ray detectability depends on whether the luminosity gain is due to higher black hole mass or a higher Eddington ratio. Based on the SED fits by \citet{Chen25}, our analysis suggests that the missing X-ray detections at high redshift are dominated by sources with high accretion rates, while high-mass, low accretion rate objects should remain detectable in X-rays. Direct comparison of the median masses in different redshift bins confirms that the \xu{} sample has on average lower black hole masses by about a factor of 2, and correspondingly a factor of 2 higher Eddington ratios. This comparison could only be done in the redshift bin $0.6<z<1.6$, where Mg\,{\sc ii}-based single-epoch estimates are currently available. The difference between both samples in other features, such as the variability amplitudes however, suggest that the \xu{} samples contains higher Eddington ratio sources at higher redshifts as well. 

The ensemble spectral properties of the \xd{} and \xu{} samples reveal two key differences:
\begin{itemize}
    \item Optical Diagnostics: The \xd{} sample exhibits stronger AGN features, including steeper continuum slopes and larger equivalent widths (EW) for broad emission lines. In contrast, the EW of forbidden and semi-forbidden lines are comparable across both samples.
    \item Absorption Features: Broad absorption lines (BALs) are found almost exclusively in the \xu{} sample. While C\,{\sc iv} broad emission lines can be suppressed by absorption, we note that the stronger (higher EW) broad emission lines in the \xd{} sample are also observed in Mg\,{\sc ii} and the Balmer series, which are largely unaffected by the BAL phenomenon. A stacking analysis of the X-ray observations in the locations of BAL and non-BAL \xu{} objects reveals that the non-BALs can be detected via the stacks while the BAL sample remains undetectable and is at least a factor of 5 weaker than the \xu{} objects detected in the stack. 
\end{itemize}
Finally, the \xd{} sample is, on average, more optically variable than the \xu{} sample and this difference is maintained across the different redshift bins. This finding is unexpected, given that the \xd{} sample is on average more luminous and the amplitude of optical/UV variability in AGN is known to anti-correlate with luminosity. This difference in variability amplitude therefore points to a connection between X-ray weakness and optical variability, either causal, i.e. less X-ray reprocessing producing lower amplitude variability, or structural, where both observables respond to a common property.

As previously discussed, the dependence of the SED on mass and accretion rate suggests that the \xu\ sample at high redshift is mainly composed of high-accretion-rate objects. The fact that optical/UV variability amplitudes on long timescales (e.g., hundreds of days) anti-correlate almost exclusively with the accretion rate, independent of black hole mass, may explain why the \xu\ sources show lower variability despite their lower average luminosities. Finally, this prevalence of high accretion rates may provide a physical explanation for why the \xu\ sample contains nearly all of the Broad Absorption Line (BAL) quasars, as these features are frequently associated with high-Eddington ratio sources \citep[e.g.][]{Petley2024}. This interpretation is further supported by the stronger blueshift of the   C\,{\sc iv} emission-line peak in the \xu{} sample.

\begin{acknowledgements}

We acknowledge support from Millennium Science Initiative Program NCN$2023\_002$ (PA, PL, SR, MLM-A, JC-M), FONDECYT Regular 1241422 (PA, PL, FEB) and 1241005 (FEB, PA), ANID-Chile BASAL project CATA FB210003 (FEB), and CAV, CIDI N. 21 U. de Valparaíso, Chile (PA), China-Chile Joint Research Fund (CCJRF2310) (MLM-A), Programa de Becas/Doctorado Nacional ANID 21263401 (JC-M). MM acknowledges support from the Spanish Ministry of Science and Innovation through the project PID2024-159201NB-C22 and also partly supported by the Spanish program Unidad de Excelencia Mar\'ia de Maeztu CEX2020-001058-M, financed by MCIN/AEI/10.13039/501100011033, and by the MaX-CSIC Excellence Award MaX4-SOMMA-ICE.

This research used data obtained with the Dark Energy Spectroscopic Instrument (DESI). DESI construction and operations is managed by the Lawrence Berkeley National Laboratory. This material is based upon work supported by the U.S. Department of Energy, Office of Science, Office of High-Energy Physics, under Contract No. DE–AC02–05CH11231, and by the National Energy Research Scientific Computing Center, a DOE Office of Science User Facility under the same contract. Additional support for DESI was provided by the U.S. National Science Foundation (NSF), Division of Astronomical Sciences under Contract No. AST-0950945 to the NSF’s National Optical-Infrared Astronomy Research Laboratory; the Science and Technology Facilities Council of the United Kingdom; the Gordon and Betty Moore Foundation; the Heising-Simons Foundation; the French Alternative Energies and Atomic Energy Commission (CEA); the National Council of Humanities, Science and Technology of Mexico (CONAHCYT); the Ministry of Science and Innovation of Spain (MICINN), and by the DESI Member Institutions: www.desi.lbl.gov/collaborating-institutions. The DESI collaboration is honored to be permitted to conduct scientific research on I’oligam Du’ag (Kitt Peak), a mountain with particular significance to the Tohono O’odham Nation. Any opinions, findings, and conclusions or recommendations expressed in this material are those of the author(s) and do not necessarily reflect the views of the U.S. National Science Foundation, the U.S. Department of Energy, or any of the listed funding agencies.

This research uses services or data provided by the SPectra Analysis and Retrievable Catalog Lab (SPARCL) and the Astro Data Lab, which are both part of the Community Science and Data Center (CSDC) Program of NSF NOIRLab. NOIRLab is operated by the Association of Universities for Research in Astronomy (AURA), Inc. under a cooperative agreement with the U.S. National Science Foundation.

This work is based on data from eROSITA, the soft X-ray instrument aboard SRG, a joint Russian-German science mission supported by the Russian Space Agency (Roskosmos), in the interests of the Russian Academy of Sciences represented by its Space Research Institute (IKI), and the Deutsches Zentrum für Luft- und Raumfahrt (DLR). The SRG spacecraft was built by Lavochkin Association (NPOL) and its subcontractors, and is operated by NPOL with support from the Max Planck Institute for Extraterrestrial Physics (MPE).
The development and construction of the eROSITA X-ray instrument was led by MPE, with contributions from the Dr. Karl Remeis Observatory Bamberg \& ECAP (FAU Erlangen-Nuernberg), the University of Hamburg Observatory, the Leibniz Institute for Astrophysics Potsdam (AIP), and the Institute for Astronomy and Astrophysics of the University of Tübingen, with the support of DLR and the Max Planck Society. The Argelander Institute for Astronomy of the University of Bonn and the Ludwig Maximilians Universität Munich also participated in the science preparation for eROSITA.
\end{acknowledgements}

\bibliographystyle{aa}
\bibliography{bibliography.bib}

\end{document}